\documentclass[twocolumn, fleqn]{doc_class}

\usepackage[super, comma, sort&compress]{natbib}

\usepackage{multibib}
\newcites{m}{References}
\bibliographystylem{bib_style}
\newcites{met}{References (continued)}
\bibliographystylemet{bib_style}
\newcites{sup}{References (continued)}
\bibliographystylesup{bib_style}

\usepackage{newtxtext, newtxmath}

\usepackage[T1]{fontenc}
\usepackage[utf8]{inputenc}

\usepackage{ae, aecompl}
\usepackage{graphicx}
\usepackage[dvipsnames, hyperref]{xcolor}
\usepackage{xspace}
\usepackage{soul}
\usepackage{booktabs}
\usepackage{adjustbox}
\usepackage{comment}

\usepackage{amsmath}
\usepackage{commath}
\usepackage{siunitx}
\usepackage{nth}

\usepackage[capitalise, noabbrev]{cleveref}
\usepackage{enumitem}
\usepackage{csvsimple}
\usepackage{multirow}
\usepackage{array}
\newcolumntype{P}[1]{>{\centering\arraybackslash}p{#1}}

\crefname{figure}{Fig.}{Figs.}
\crefname{equation}{equation}{equations}

\usepackage[leftmargin=0pt,rightmargin=0pt,skipabove=2pt,skipbelow=2pt,innerleftmargin=5pt,innerrightmargin=5pt,innertopmargin=5pt,innerbottommargin=5pt,linewidth=0pt,innerlinewidth=0pt,middlelinewidth=0pt,outerlinewidth=0pt,roundcorner=0pt]{mdframed}

\newcommand{\orcidsymb}[2]{#1\href{http://orcid.org/#2}{\adjustbox{trim={-.15\width} {0\height} {-.15\width} {0\height},clip}{\includegraphics[height=10pt]{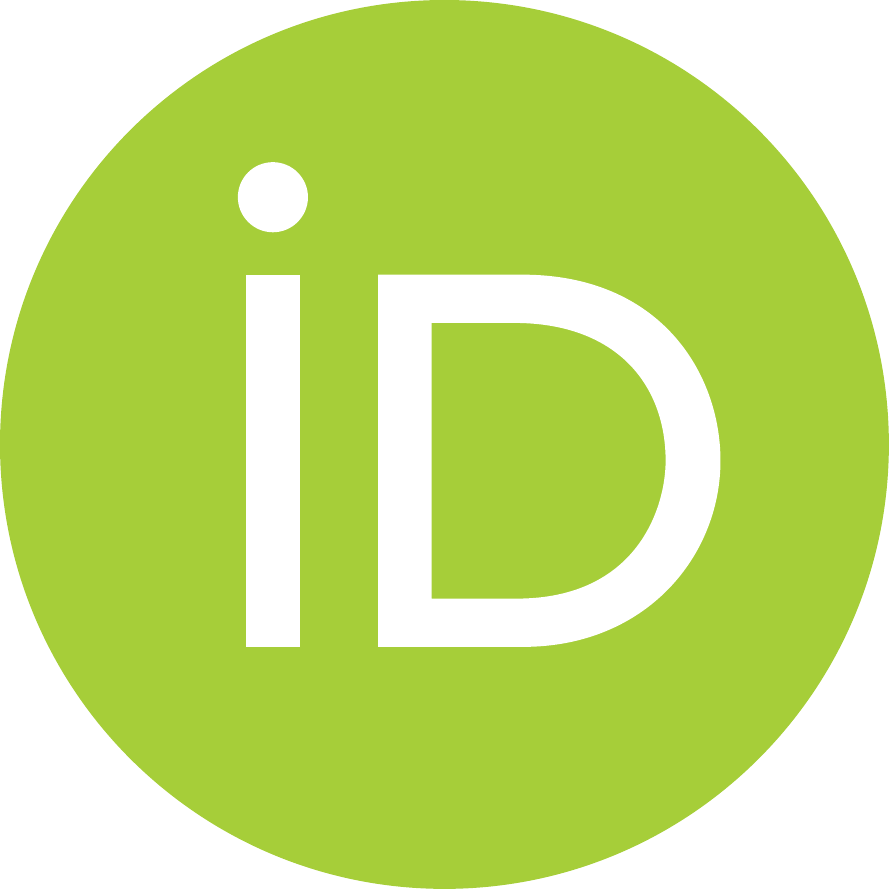}}}}

\newcommand{\Lymana}{{Lyman-\ensuremath{\upalpha}}\xspace}
\newcommand{\Lymanatext}{Lyman-α}
\newcommand{\Lya}{{Ly\ensuremath{\upalpha}}\xspace}

\newcommand{\HI}{\hbox{H\,{\sc i}}\xspace}
\newcommand{\HII}{\hbox{H\,{\sc ii}}\xspace}
\newcommand{\CIV}{\hbox{C\,{\sc iv}}\xspace}

\newcommand{\HeII}{\hbox{He\,{\sc ii}}\xspace}

\newcommand{\CIIIs}{\hbox{C\,{\sc iii}]}\xspace}
\newcommand{\CIIIf}{\hbox{[C\,{\sc iii}]}\xspace}

\newcommand{\OII}{\hbox{[O\,{\sc ii}]}\xspace}
\newcommand{\OIII}{\hbox{[O\,{\sc iii}]}\xspace}
\newcommand{\OIIIs}{\hbox{O\,{\sc iii}]}\xspace}

\newcommand{\NII}{\hbox{[N\,{\sc ii}]}\xspace}
\newcommand{\NIII}{\hbox{N\,{\sc iii}]}\xspace}
\newcommand{\NIV}{\hbox{N\,{\sc iv}]}\xspace}

\newcommand{\Halpha}{\ensuremath{\mathrm{H}\upalpha}\xspace}
\newcommand{\Hbeta}{\ensuremath{\mathrm{H}\upbeta}\xspace}

\newcommand{\Paalpha}{\ensuremath{\mathrm{Pa}\upalpha}\xspace}

\newcommand{\PopulationIII}{\hbox{Population\,{\sc iii}}\xspace}
\newcommand{\PopIII}{\hbox{Pop\,{\sc iii}}\xspace}

\newcommand{\JGSzthirteenLA}{JADES-GS-z13-1-LA\xspace}

\hypersetup{
    unicode=true,
    final=true,
    plainpages=false,
    pdfstartview=FitV,
    pdftoolbar=false,
    pdfmenubar=true,
    bookmarksopen=true,
    bookmarksnumbered=true,
    breaklinks=true,
    linktocpage,
    colorlinks=true,
    linkcolor=Blue,
    urlcolor=Blue,
    citecolor=Blue,
    anchorcolor=green
}
\AtBeginDocument{
    \hypersetup{
        pdftitle={Witnessing the onset of reionisation via Lyman-α emission at redshift 13},
        pdfauthor={J. Witstok, P. Jakobsen et al.},
        pdfsubject={},
        pdfkeywords={{galaxies: high-redshift} -- {dark ages, reionization, first stars} -- {methods: observational} -- {techniques: spectroscopic}}
    }
}

\usepackage{paralist}
\makeatletter
    \def\@biblabel#1{\@ifnotempty{#1}{#1}}
    \def\NAT@anchor#1#2{
        \hfilneg\hyper@natanchorstart{#1\@extra@b@citeb}
        #2.
        \hyper@natanchorend
    }
\makeatother

\renewenvironment{thebibliography}[1]{%
    \newpage
    \section*{\normalsize \refname}
    \setlength{\parindent}{0pt}
    \inparaenum
}{\endinparaenum}

\DeclareRobustCommand{\VAN}[3]{#2}
\let\VANthebibliography\thebibliography
\def\thebibliography{\DeclareRobustCommand{\VAN}[3]{##3}\VANthebibliography}

\title[Witnessing the onset of reionisation via \texorpdfstring{\Lymana}{\Lymanatext} emission at redshift 13]{Witnessing the onset of reionisation via \texorpdfstring{\Lymana}{\Lymanatext} emission at redshift 13}

\author[]{{\orcidsymb{Joris Witstok}{0000-0002-7595-121X}$^{\hyperlink{inst:Kavli}{1}, \hyperlink{inst:Cav}{2}, \hyperlink{inst:DAWN}{3}, \hyperlink{inst:NBI}{4}}$\thanks{E-mail: \href{mailto:joris.witstok@nbi.ku.dk}{joris.witstok@nbi.ku.dk}}, \orcidsymb{Peter Jakobsen}{0000-0002-6780-2441}$^{\hyperlink{inst:DAWN}{3}, \hyperlink{inst:NBI}{4}}$, \orcidsymb{Roberto Maiolino}{0000-0002-4985-3819}$^{\hyperlink{inst:Kavli}{1}, \hyperlink{inst:Cav}{2}, \hyperlink{inst:UCL}{5}}$, \orcidsymb{Jakob M. Helton}{0000-0003-4337-6211}$^{\hyperlink{inst:Steward}{6}}$,
    }
    \newauthor{\orcidsymb{Benjamin D.\ Johnson}{0000-0002-9280-7594}$^{\hyperlink{inst:CfA}{7}}$, \orcidsymb{Brant E.\ Robertson}{0000-0002-4271-0364}$^{\hyperlink{inst:UCSC}{8}}$, \orcidsymb{Sandro Tacchella}{0000-0002-8224-4505}$^{\hyperlink{inst:Kavli}{1}, \hyperlink{inst:Cav}{2}}$, \orcidsymb{Alex J.\ Cameron}{0000-0002-0450-7306}$^{\hyperlink{inst:Oxford}{9}}$,
    }
    \newauthor{\orcidsymb{Renske Smit}{0000-0001-8034-7802}$^{\hyperlink{inst:LJMU}{10}}$, \orcidsymb{Andrew J.\ Bunker}{0000-0002-8651-9879}$^{\hyperlink{inst:Oxford}{9}}$, \orcidsymb{Aayush Saxena}{0000-0001-5333-9970}$^{\hyperlink{inst:Oxford}{9}, \hyperlink{inst:UCL}{5}}$, \orcidsymb{Fengwu Sun}{0000-0002-4622-6617}$^{\hyperlink{inst:Steward}{6}, \hyperlink{inst:CfA}{7}}$, \orcidsymb{Stacey Alberts}{0000-0002-8909-8782}$^{\hyperlink{inst:Steward}{6}}$,
    }
    \newauthor{\orcidsymb{Santiago Arribas}{0000-0001-7997-1640}$^{\hyperlink{inst:CAB}{11}}$, \orcidsymb{William M.\ Baker}{0000-0003-0215-1104}$^{\hyperlink{inst:Kavli}{1}, \hyperlink{inst:Cav}{2}}$, \orcidsymb{Rachana Bhatawdekar}{0000-0003-0883-2226}$^{\hyperlink{inst:ESAC}{12}}$, \orcidsymb{Kristan Boyett}{0000-0003-4109-304X}$^{\hyperlink{inst:Oxford}{9}}$,
    }
    \newauthor{\orcidsymb{Phillip A.\ Cargile}{0000-0002-1617-8917}$^{\hyperlink{inst:CfA}{7}}$, \orcidsymb{Stefano Carniani}{0000-0002-6719-380X}$^{\hyperlink{inst:SNS}{13}}$, \orcidsymb{St\'ephane Charlot}{0000-0003-3458-2275}$^{\hyperlink{inst:IAP}{14}}$, \orcidsymb{Jacopo Chevallard}{0000-0002-7636-0534}$^{\hyperlink{inst:Oxford}{9}}$,
    }
    \newauthor{\orcidsymb{Mirko Curti}{0000-0002-2678-2560}$^{\hyperlink{inst:ESO}{15}}$, \orcidsymb{Emma Curtis-Lake}{0000-0002-9551-0534}$^{\hyperlink{inst:Herts}{16}}$, \orcidsymb{Francesco D'Eugenio}{0000-0003-2388-8172}$^{\hyperlink{inst:Kavli}{1}, \hyperlink{inst:Cav}{2}, \hyperlink{inst:INAF}{17}}$, \orcidsymb{Daniel J.\ Eisenstein}{0000-0002-2929-3121}$^{\hyperlink{inst:CfA}{7}}$,
    }
    \newauthor{\orcidsymb{Kevin N.\ Hainline}{0000-0003-4565-8239}$^{\hyperlink{inst:Steward}{6}}$, \orcidsymb{Gareth C.\ Jones}{0000-0002-0267-9024}$^{\hyperlink{inst:Oxford}{9}, \hyperlink{inst:Kavli}{1}, \hyperlink{inst:Cav}{2}}$, \orcidsymb{Nimisha Kumari}{0000-0002-5320-2568}$^{\hyperlink{inst:AURA}{18}}$, \orcidsymb{Michael V.\ Maseda}{0000-0003-0695-4414}$^{\hyperlink{inst:Wisconsin}{19}}$,
    }
    \newauthor{\orcidsymb{Pablo G.\ P\'erez-Gonz\'alez}{0000-0003-4528-5639}$^{\hyperlink{inst:CAB}{11}}$, \orcidsymb{Pierluigi Rinaldi}{0000-0002-5104-8245}$^{\hyperlink{inst:Steward}{6}}$, \orcidsymb{Jan Scholtz}{0000-0001-6010-6809}$^{\hyperlink{inst:Kavli}{1}, \hyperlink{inst:Cav}{2}}$, \orcidsymb{Hannah \"Ubler}{0000-0003-4891-0794}$^{\hyperlink{inst:MPE}{20}}$,
    }
    \newauthor{\orcidsymb{Christina C.\ Williams}{0000-0003-2919-7495}$^{\hyperlink{inst:NOIRLab}{21}}$, \orcidsymb{Christopher N.\ A.\ Willmer}{0000-0001-9262-9997}$^{\hyperlink{inst:Steward}{6}}$, \orcidsymb{Chris Willott}{0000-0002-4201-7367}$^{\hyperlink{inst:NRC}{22}}$, and \orcidsymb{Yongda Zhu}{0000-0003-3307-7525}$^{\hyperlink{inst:Steward}{6}}$
    }
    \\
    \\
    {\normalsize Affiliations are listed at the end of the manuscript.}
}

\pubyear{2025}

\begin{document}
\maketitle

\begin{abstract}
    \begin{mdframed}[backgroundcolor=black!5]
        Cosmic Reionisation commenced when ultraviolet (UV) radiation produced in the first galaxies began illuminating the cold, neutral gas that filled the primordial Universe\citem{2018PhR...780....1D, 2022ARA&A..60..121R}. Recent James Webb Space Telescope (JWST) observations have shown that surprisingly UV-bright galaxies were in place beyond redshift $z = 14$, when the Universe was less than $300 \, \mathrm{Myr}$ old\citem{2023NatAs...7..622C, 2023Natur.622..707A, 2024Natur.633..318C}. Smooth turnovers of their UV continua have been interpreted as damping-wing absorption of \Lymana (\Lya), the principal hydrogen transition\citem{2024A&A...689A.152D, 2024ApJ...973....8H, 2024Sci...384..890H, 2024ApJ...976..160H}. However, spectral signatures encoding crucial properties of these sources, such as their emergent radiation field, largely remain elusive. Here we report spectroscopy from the JWST Advanced Deep Extragalactic Survey (JADES\citem{2023arXiv230602465E}) of a galaxy at redshift $z = 13.0$ that reveal a singular, bright emission line unambiguously identified as \Lya, in addition to a smooth turnover. We observe an equivalent width of $\text{EW}_\text{\Lya} > 40 \, \Angstrom$ (rest frame), previously only seen at $z < 9$ where the intervening intergalactic medium (IGM) becomes increasingly ionised\citem{2025MNRAS.536...27W}. Together with an extremely blue UV continuum, the unexpected \Lya emission indicates the galaxy is a prolific producer and leaker of ionising photons. This suggests massive, hot stars or an active galactic nucleus (AGN) have created an early reionised region to prevent complete extinction of \Lya, thus shedding new light on the nature of the earliest galaxies and the onset of Reionisation only $330 \, \mathrm{Myr}$ after the Big Bang.
    \end{mdframed}
\end{abstract}
\nokeywords


\begin{figure*}
	\centering
	\includegraphics[width=\linewidth]{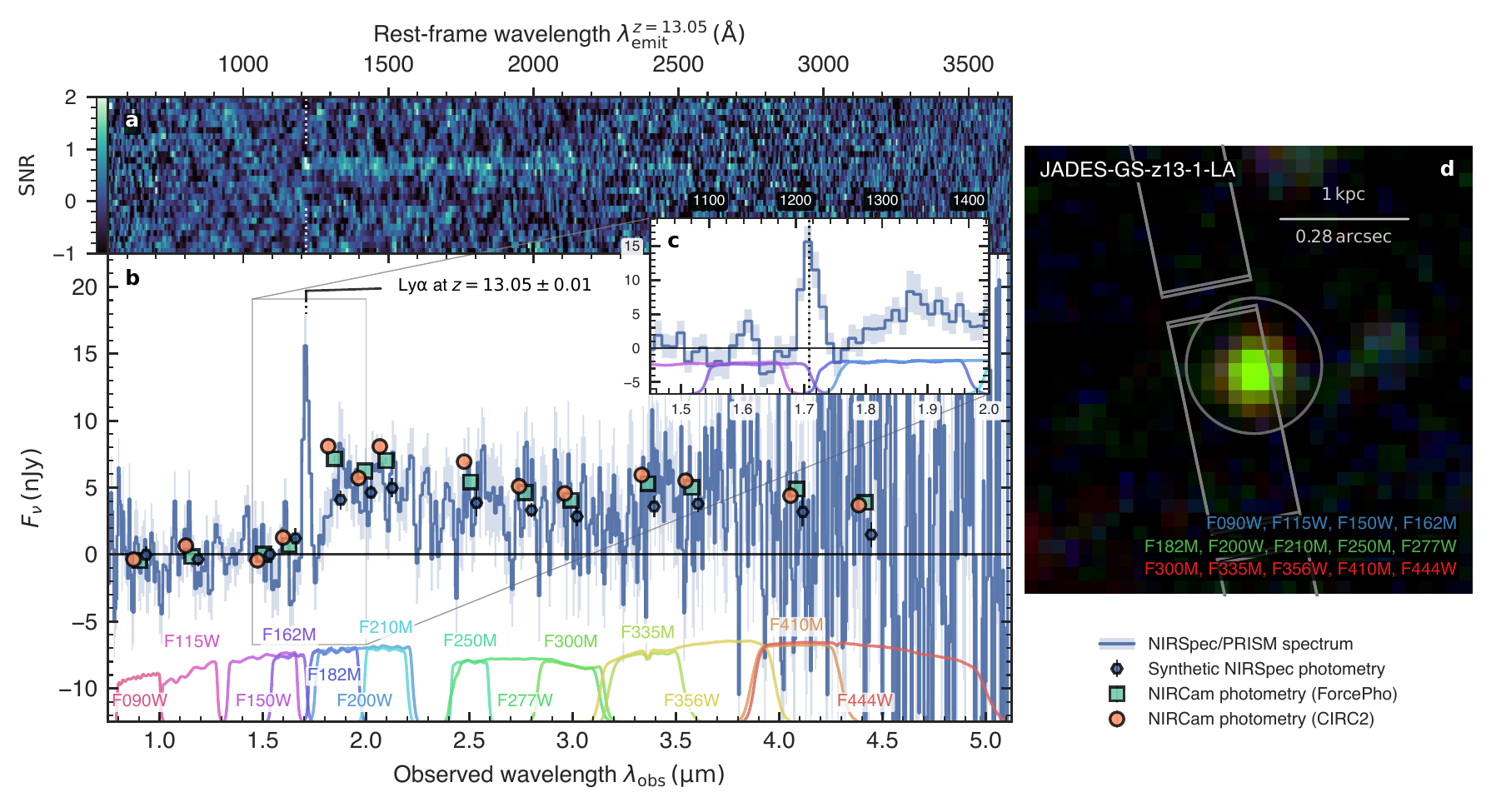}
	\caption{\textbf{NIRCam and NIRSpec/PRISM observations of \JGSzthirteenLA.} \textbf{a}, Two-dimensional SNR map of the PRISM spectrum (not used for extraction of the one-dimensional spectrum; see Methods for details). \textbf{b}, One-dimensional sigma-clipped PRISM spectrum (uncorrected for additional path losses; Methods) and photometric measurements (slightly offset in wavelength for visualisation) according to the legend in the bottom right. Synthetic photometry is obtained by convolving the spectrum with the filter transmission curves shown at the bottom. Shading and error bars represent $1 \sigma$ uncertainty. \textbf{c}, Zoom-in on the emission line at $1.7 \, \mathrm{\upmu m}$, which falls precisely in between the F162M and F182M medium-band filters. \textbf{d}, False-colour image of \JGSzthirteenLA constructed by stacking NIRCam filters for each colour channel as annotated. The placement of the NIRSpec micro-shutters, nearly identical across the two visits, are shown in grey, as is the circular $0.3\arcsec$-diameter extraction aperture for the CIRC2 photometry. A physical scale of $1 \, \mathrm{kpc}$ ($0.28\arcsec$ at $z = 13.05$) is indicated.
	}
	\label{fig:NIRCam_NIRSpec}
\end{figure*}

\noindent Using the Near-Infrared Camera (NIRCam\citem{2023PASP..135b8001R}) and Mid-Infrared Instrument (MIRI\citem{2015PASP..127..584R}) on board JWST, we obtained deep imaging as part of the JADES and JADES Origins Field (JOF\citem{2023arXiv231012340E}) programmes. A careful search for high-redshift galaxy candidates exploiting the 14-band NIRCam coverage\citem{2024ApJ...964...71H, 2024ApJ...970...31R} led to the identification of JADES-GS+53.06475-27.89024 (\JGSzthirteenLA hereafter) as the most robust redshift $z \gtrsim 11.5$ photometric candidate in the JOF based on its blue colour and clear `dropout' signature, confidently rejecting a brown dwarf solution. Since the discontinuity strength ($>20\times$ in flux between the NIRCam F150W and F200W filters) further rules out a Balmer break due to an evolved stellar population at much lower redshift, the photometry strongly favours a solution at $z \approx 13$ where \Lya, the $2p \rightarrow 1s$ electronic transition of hydrogen, is shifted to $1.7 \, \mathrm{\upmu m}$ in the observed frame and any photons emitted at shorter wavelengths are completely absorbed by neutral hydrogen (\HI) in the intervening IGM.

Follow-up spectroscopy of \JGSzthirteenLA was obtained as part of JADES with the JWST Near-Infrared Spectrograph (NIRSpec\citem{2022A&A...661A..80J}), principally in PRISM mode (exposure time of $18.7 \, \mathrm{h}$), covering wavelengths from $0.6 \, \mathrm{\upmu m}$ up to $5.3 \, \mathrm{\upmu m}$ at low resolution ($R \approx 100$). As shown in \cref{fig:NIRCam_NIRSpec}, the resulting spectrum unequivocally confirms the redshift to be $z \approx 13.0$ (Methods) even if the break is smooth rather than sharp, which indeed is expected for sources embedded in a highly neutral IGM due to \Lya damping-wing absorption\citem{1998ApJ...501...15M}, as has been seen directly in quasar spectra\citem{2018Natur.553..473B}. Spectra of $z \gtrsim 9$ galaxies recently discovered by JWST have also hinted at the existence of IGM damping wings\citem{2023NatAs...7..622C, 2024ApJ...973....8H}, although many cases have been observed to far exceed pure IGM absorption, which has been ascribed to local damped \Lya (DLA) absorbing systems (column densities of $N_\text{\HI} > 10^{20.3} \, \mathrm{cm^{-2}}$; ref. \citem{2005ARA&A..43..861W}) interpreted as pockets of dense, neutral gas within or near the galaxy\citem{2024A&A...689A.152D, 2024Sci...384..890H, 2024ApJ...976..160H, 2024arXiv240920533C}.

Remarkably, unlike any other $z > 10$ galaxies confirmed by JWST\citem{2023NatAs...7..622C, 2023A&A...677A..88B, 2023Natur.622..707A, 2023ApJ...957L..34W, 2024A&A...689A.152D, 2024ApJ...960...56H, 2024Natur.633..318C}, the PRISM spectrum additionally reveals a bright emission line detected at high signal-to-noise ratio ($\text{SNR} = 6.4$) and consistently across the two independent visits (Methods). Located at the blue edge of the spectral break, it is observed at $\lambda_\text{obs} = 1.7084 \pm 0.0014 \, \mathrm{\upmu m}$ and while the continuum directly underneath is not detected, we can conservatively place a lower limit on the rest-frame EW of $>40 \, \Angstrom$. The only viable explanation, considering the clear break and the absence of nearby foreground sources and any other lines (Methods), is to identify the line as \Lya at a redshift of $z_\text{\Lya} = 13.05 \pm 0.01$. However, due to the resonant nature of \Lya, we note the systemic redshift is likely slightly lower.

If not arising from collisional excitation, expected to be subdominant even at interstellar medium (ISM) densities of $n \approx 10^4 \, \mathrm{cm^{-3}}$ (ref. \citem{2010A&A...523A..64R}), this immediately implies that \JGSzthirteenLA produces a substantial number of ionising Lyman-continuum (LyC) photons as quantified by the production efficiency, for which we find a robust lower limit of $\xi_\text{ion} \gtrsim 10^{25.1} \, \mathrm{Hz \, erg^{-1}}$ (Methods). While already close to the canonical value required for star-forming galaxies to complete reionisation\citem{2015ApJ...802L..19R}, this value increases considerably if any \Lya photons are absorbed within the galaxy or scattered out of our line of sight in the IGM. This should be a major effect at $z = 13$ as the Universe is still highly neutral\citem{2018Natur.553..473B, 2024ApJ...973....8H}, even if a local ionised `bubble' around the galaxy facilitates the transmission of \Lya photons\citem{2025MNRAS.536...27W}. Note that while photon diffusion via resonant scattering off neutral gas in the IGM is predicted to result in extended \Lya halos around galaxies before reionisation\citem{1999ApJ...524..527L}, such diffuse emission cannot explain the observed line properties. From non-detections in our medium-resolution spectra, although less sensitive than the PRISM, we do however infer the line is likely broadened spectrally (Methods).

\begin{figure*}
	\centering
	\includegraphics[width=\linewidth]{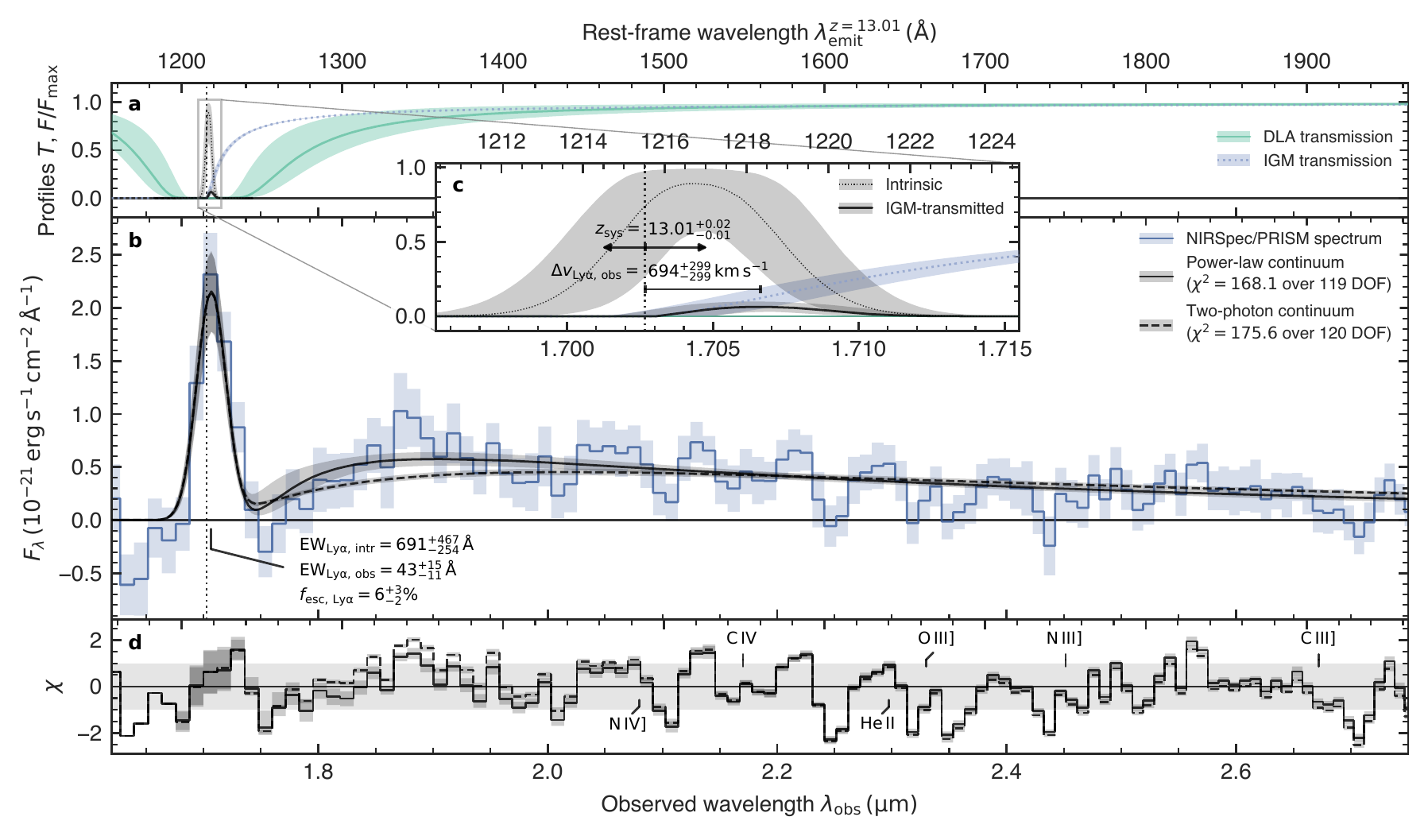}
	\caption{\textbf{Model of NIRSpec/PRISM observations of \JGSzthirteenLA.} \textbf{a}, Model curves for the IGM and DLA transmission $T$ (according to the legend on the right) and normalised \Lya line profiles (cf. panel~c). \textbf{b}, A blue line shows the sigma-clipped PRISM spectrum corrected for path losses (Methods). Model spectra with a power-law continuum, attenuated by DLA absorption, and a pure $2\gamma$ continuum are shown by the black solid and dashed lines, respectively. The legend shows their $\chi^2$ goodness-of-fit statistics compared to the degrees of freedom (DOF; Methods). The intrinsic and observed \Lya EWs (relative to an unattenuated power-law continuum), and their ratio (the escape fraction), are annotated. A similar annotation indicates upper limits on the \HeII line flux and EW ($3\sigma$; Methods). \textbf{c}, Zoom-in on the intrinsic (dotted black) and IGM-transmitted (solid black) \Lya line profiles. The vertical black dotted line shows the median systemic \Lya redshift in the default model (Methods), differing from the \Lya redshift by the observed velocity offset $\Delta v_\text{\Lya, obs}$. \textbf{d}, For the two different models, $\chi$ represents the residuals normalised by the observational uncertainty of a single wavelength bin (diagonal elements of the covariance matrix). The location of other rest-frame UV lines are indicated, though none are significantly detected (Methods). Shading represents $1\sigma$ uncertainty on all lines.
	}
	\label{fig:NIRSpec_model}
\end{figure*}

Fitting a variety of standard stellar population synthesis (SPS) models to the observed spectral energy distribution (SED) of \JGSzthirteenLA yields a young ($10$-$20 \, \mathrm{Myr}$) and metal-poor ($<2\%$ Solar) stellar population, with little to no dust obscuration (Supplementary information). However, commonly used SED fitting codes do not have the capability to model the peculiar coexistence of \Lya emission together with a smooth spectral turnover. To better understand its origin in \JGSzthirteenLA, we therefore performed detailed spectral modelling where we take into account potential absorption by DLA absorbers, transmission through a neutral, mean-density IGM with a local ionised bubble, and instrumental effects such as path losses and the line spread function (LSF). For our fiducial model, we opt for a power-law continuum that offers the flexibility to recreate the steep UV slope, which from the NIRCam and NIRSpec data we consistently measure to be $\beta_\text{UV} \lesssim -2.7$ (Methods). However, we also considered the inclusion of nebular continuum, since the two-photon ($2\gamma$) continuum in lower-redshift galaxies has been suggested\citem{2004AJ....127.3146H, 2024MNRAS.534..523C} as the potential origin of a UV turnover and \Lya emission qualitatively similar to \JGSzthirteenLA. Best-fitting models with a pure power-law and $2\gamma$ continuum are shown in \cref{fig:NIRSpec_model}.

Regardless of the choice of continuum, our model indicates that across a range of reasonable emergent \Lya profiles, approximately $5$-$10\%$ of flux may be transmitted through the IGM, implying an intrinsic \Lya luminosity of $L_\text{\Lya} \approx 2 \times 10^{43} \, \mathrm{erg \, s^{-1}}$. Here, we allow for a non-zero LyC escape fraction causing a local ionised bubble with radius $R_\text{ion} \approx 0.2 \, \text{physical Mpc}$ (pMpc) to form within an otherwise neutral IGM, without which the required luminosity would triple, a scenario disfavoured by the non-detection in the MIRI/F770W filter containing \Hbeta (Methods). Still, we find the models consistently require $\xi_\text{ion} \approx 10^{26.5} \, \mathrm{Hz \, erg^{-1}}$, either to create the transmission-enhancing bubble or boost the intrinsic luminosity. For any appreciable IGM transmission, the observed \Lya peak should fall substantially redwards ($\Delta v_\text{\Lya, obs} \gtrsim 500 \, \mathrm{km \, s^{-1}}$) of the systemic redshift\citem{2023A&A...677A..88B, 2024A&A...687A.283S}, which we therefore infer to be $z_\text{sys} = 13.01_{-0.01}^{+0.02}$.

For standard stellar models, the remarkably high $\xi_\text{ion}$ is untenable\citem{2022ApJ...941..153T, 2022ARA&A..60..455E} under common initial mass functions (IMFs). Since $\xi_\text{ion}$ is directly sensitive to the hottest stars, its extreme value may be ascribed to an extension of the IMF to very massive stars\citem{2020MNRAS.493.5120M, 2023A&A...678A.173V}. The high average ionising-photon energy of a $T = 10^5 \, \mathrm{K}$ blackbody moreover yields a $2\times$ higher ratio of \Lya to LyC photons than standard case-B recombination\citem{2010A&A...523A..64R}, thereby bringing the true $\xi_\text{ion}$ more closely in agreement with the theoretical stellar maximum\citem{2024arXiv240712122S}. One particularly intriguing class of objects predicted to radiate up to $40\%$ of their bolometric luminosity as \Lya are entirely metal-free \PopulationIII (\PopIII) stars\citem{2002A&A...382...28S, 2023MNRAS.524..351K, 2023MNRAS.525.5328T} thought to reach significantly higher masses and effective temperatures than subsequent metal-enriched stellar populations. However, the absolute UV magnitude of \JGSzthirteenLA, $M_\text{UV} \approx -18.7 \, \mathrm{mag}$, would require a stellar mass of $M_* \approx 10^6 \, \mathrm{M_\odot}$ as a pure \PopIII system, somewhat higher than typical predictions\citem{2011ARA&A..49..373B}. Furthermore, the absence of strong $\HeII \, \lambda \, 1640 \, \Angstrom$ emission (Methods) may argue against the \PopIII scenario\citem{2022MNRAS.513.5134N}, though its strength rapidly evolves a few $\mathrm{Myr}$ after a star formation burst\citem{2023MNRAS.524..351K, 2023MNRAS.525.5328T}.

The presence of extraordinarily hot stars ($T_\text{eff} > 10^5 \, \mathrm{K}$) required to explain such high $\xi_\text{ion}$ could naturally lead relatively low-density gas ($n \lesssim 10^4 \, \mathrm{cm^{-3}}$) to emit a prominent nebular continuum with a UV turnover\citem{2024MNRAS.534..523C, 2024arXiv240803189K}. However, we find that compared to the pure $2\gamma$ continuum, which only becomes further reddened by free-bound continuum emission at higher densities, the current data are better reproduced by a steep power law (\cref{fig:NIRSpec_model}). A scenario where \Lya emission is produced together with $2\gamma$ continuum as cooling radiation via collisional excitation in the dense core of a collapsing cloud\citem{2009ApJ...690...82D} is therefore also disfavoured. The extremely blue UV continuum ($\beta_\text{UV} \lesssim -2.7$) consistently leads our models to prefer near-unity LyC escape fraction to reproduce the blue SED of \JGSzthirteenLA, even with an IMF extending to $300 \, \mathrm{M_\odot}$ (Methods). Moreover, recent stellar models show\citem{2022A&A...659A.163M} the effective temperatures of very massive stars stagnate beyond $100 \, \mathrm{M_\odot}$, suggesting a high LyC escape fraction remains necessary. While this would corroborate the suggestion that \JGSzthirteenLA is located inside an ionised bubble and could suppress \HeII, it still leaves the UV turnover to be explained.
\begin{figure*}
	\centering
	\includegraphics[width=\linewidth]{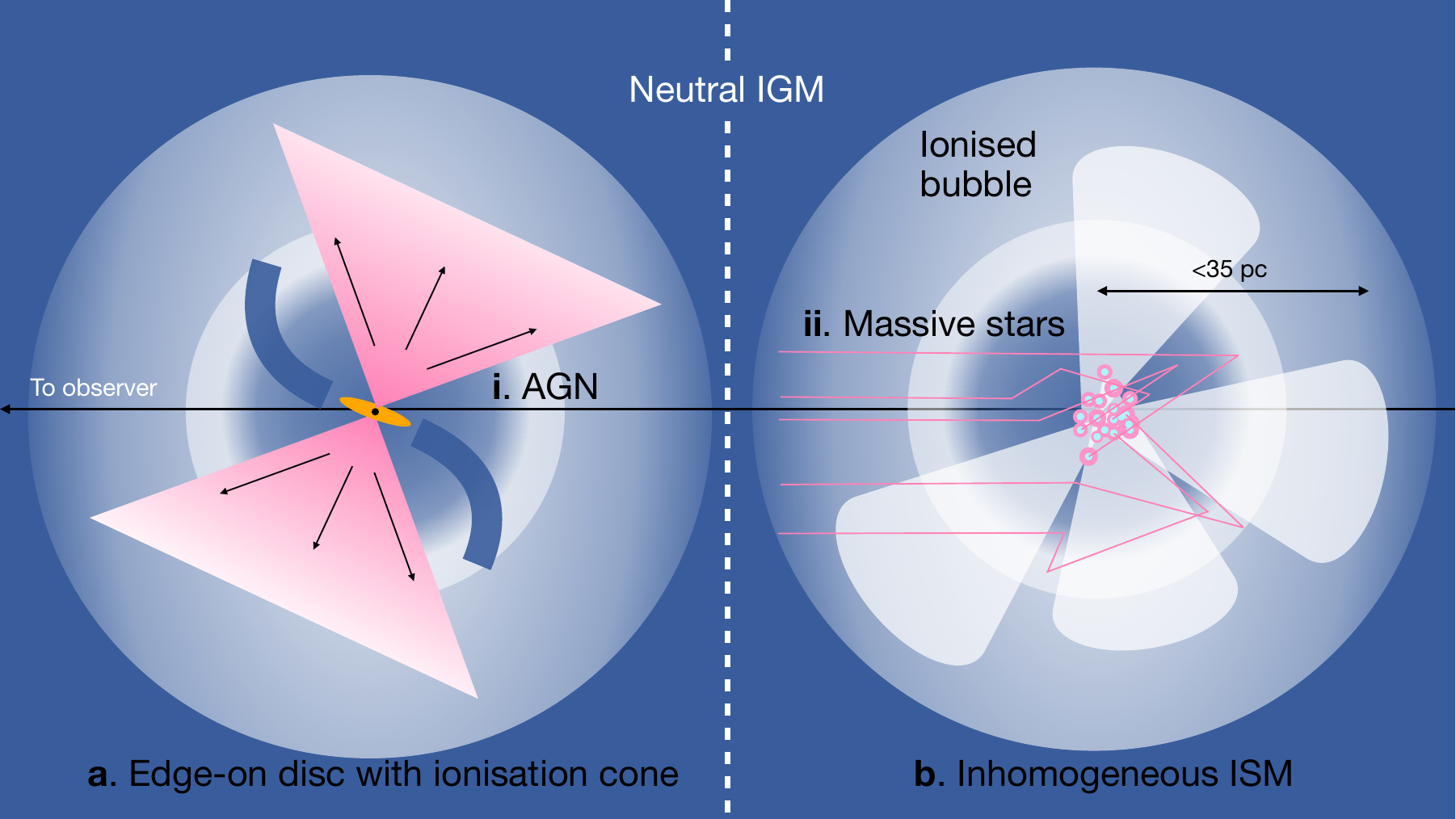}
	\caption{\textbf{Schematic of production, escape, and absorption of \Lya in \JGSzthirteenLA.} \Lya emission is indicated in pink, whereas dark blue shows \HI gas. We identify two potential explanations each for the source of emission (i and ii) and modes of \Lya modulation (a and b). An extended disc of neutral gas seen in edge-on orientation (a) may cause DLA absorption of the continuum source, while an ionisation cone perpendicular to the disc plane allows \Lya photons to escape. Under this escape mechanism, the source of the \Lya emission may be interchanged from an AGN (i) to a nuclear starburst (ii). Alternatively, if neutral gas in the ISM is inhomogeneously distributed (b), resonant scattering could allow \Lya to diffuse outwards while the central source remains obscured by \HI gas, as seen in local, compact star-forming galaxies (see text for details).
	}
	\label{fig:Schematic}
\end{figure*}

Instead, the spectrum of \JGSzthirteenLA therefore appears to necessitate significant DLA absorption ($N_\text{\HI} \approx 10^{22.8} \, \mathrm{cm^{-2}}$ for the power-law continuum), as seen in several $z > 10$ galaxies\citem{2024A&A...689A.152D, 2024Sci...384..890H, 2024ApJ...976..160H, 2024arXiv240920533C}. If the DLA absorber were co-located with the galaxy, a specific geometry is required to simultaneously accommodate the escape of \Lya and potentially LyC. As illustrated in \cref{fig:Schematic}, an inhomogeneous ISM or an edge-on disc and associated ionisation cone may cause DLA absorption in compact continuum sources which is circumvented by \Lya emission\citem{2024arXiv240402194T}. Especially in the absence of dust, \Lya emission could escape through resonant scattering while also becoming broadened in velocity space, consistent with observations. Empirically, \Lya emission superimposed on DLA absorption has not only been reported for nearby UV-bright star-forming galaxies where it has been interpreted as a sign of ISM inhomogeneity\citem{2023ApJ...956...39H}, but also in the case of AGN\citem{2024MNRAS.532.4703W}.

Indeed, an accreting supermassive black hole (SMBH) may offer a comprehensive alternative explanation for the observed properties of \JGSzthirteenLA. Effectively unresolved by NIRCam, its half-light radius of $\lesssim 35 \, \mathrm{pc}$ (Methods) is smaller than most $z > 10$ galaxies\citem{2023NatAs...7..622C, 2023A&A...677A..88B, 2023ApJ...957L..34W, 2024Natur.633..318C}. AGN have been observed\citem{2019MNRAS.484.2575T} to reach UV slopes significantly steeper than the standard thin-disc model\citem{1973A&A....24..337S} with $\beta_\text{UV} = 7/3 \approx -2.33$, as expected for a truncated accretion disc. They are also found\citem{2018A&A...613A..44G} to have high LyC escape fractions, and the broad \Lya line could be linked to AGN-driven outflows or a broad line region. Constraints on the currently undetected \HeII and other UV lines (Methods) are consistent with model predictions for metal-poor AGN\citem{2022MNRAS.513.5134N}, altogether rendering \JGSzthirteenLA a viable candidate.

Whether the \Lya emission of \JGSzthirteenLA originates in stars or a SMBH, it reveals the rather extreme character of one of the earliest galaxies known, despite having been found in a modest survey area\citem{2024ApJ...970...31R} probing a comoving volume of $\num{50000} \, \mathrm{Mpc^3}$ between $z = 11$ and $z = 15$. At just $330 \, \mathrm{Myr}$ after the Big Bang, the likely presence of a reionised region around this relatively UV-faint source readily constrains the timeline of cosmic reionisation, favouring an early and gradual process driven (initially) by low-mass galaxies\citem{2025MNRAS.tmpL...1Q}. Furthermore, it provides tangible evidence for the Wouthuysen-Field coupling of the spin temperature of neutral hydrogen to that of the gas via the emission of \Lya photons, the global evolution of which is anticipated to be uncovered soon by \HI $21 \, \mathrm{cm}$ experiments\citem{2022NatAs...6..984D} to provide a complementary view of Cosmic Dawn.
\bibliographym{GS-z13-1-LA}

\clearpage
\makeatletter
\renewcommand{\fnum@figure}{Extended Data Fig. \thefigure}
\renewcommand{\fnum@table}{Extended Data Table \thetable}
\makeatother
\setcounter{figure}{0}
\setcounter{table}{0}
\setcounter{footnote}{0}



\begin{figure*}
	\centering
	\includegraphics[width=0.98\linewidth]{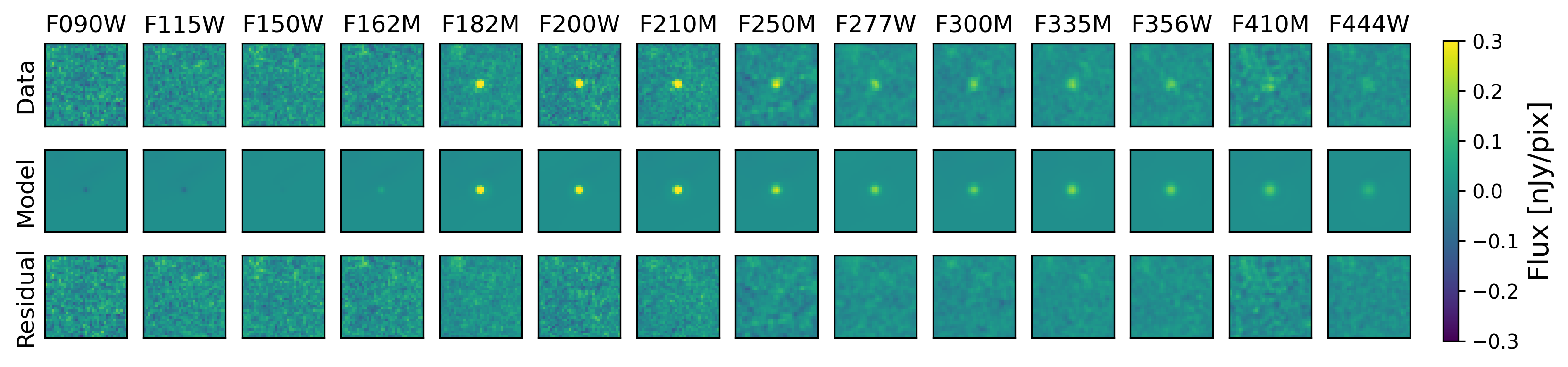}
	\caption{\textbf{\textsc{forcepho} modelling of \JGSzthirteenLA.} The top row shows $\sim 1\arcsec \times 1\arcsec$ cutouts of the observed data (scaled according to the colourbar shown on the right) around \JGSzthirteenLA in each of the $14$ available NIRCam filters, as annotated at the top of each column. The PSF-convolved \textsc{forcepho} model (see Photometric measurements) is shown in the middle row. The bottom row shows residuals between data and model are consistent with pure noise, indicating the model provides a good fit to the data. Note that while the \textsc{forcepho} fits are performed on $>400$ separate exposures, they are mosaiced together here to visualise the data and residuals.
	}
	\label{fig:ForcePho_fit}
\end{figure*}
\begingroup
    \setlength{\tabcolsep}{6pt} 
    \renewcommand{\arraystretch}{1.25} 
    \begin{table*}
        \centering
        \caption
        {\textbf{Photometry and UV-continuum properties of \JGSzthirteenLA.}}
        \label{tab:Photometry}
        \begin{tabular}{lccccc}
    \toprule
    Quantity & Instrument & Filter (set) & ForcePho & CIRC2 & Synthetic (NIRSpec)
    \\
    \midrule
    \multirow{22}{*}{Flux density $F_\nu \, (\mathrm{nJy})$} & HST/ACS & F435W & \dots & $2.37 \pm 3.81$ & \dots
    \\
    & & F606W & \dots & $-0.80 \pm 3.02$ & \dots
    \\
    & & F775W & \dots & $-3.00 \pm 4.34$ & \dots
    \\
    & & F814W & \dots & $6.11 \pm 3.79$ & \dots
    \\
    & & F850LP & \dots & $-2.25 \pm 7.35$ & \dots
    \\
    & HST/WFC3 & F125W & \dots & $-34.6 \pm 15.9$ & \dots
    \\
    & & F160W & \dots & $-12.1 \pm 22.0$ & \dots
    \\
    & JWST/NIRCam & F090W & $-0.448 \pm 0.190$ & $-0.372 \pm 0.609$ & $-0.023 \pm 0.641$
    \\
    & & F115W & $-0.157 \pm 0.163$ & $0.664 \pm 0.496$ & $-0.354 \pm 0.551$
    \\
    & & F150W & $0.023 \pm 0.183$ & $-0.430 \pm 0.480$ & $0.013 \pm 0.515$
    \\
    & & F162M & $0.696 \pm 0.211$ & $1.270 \pm 0.566$ & $1.212 \pm 0.780$
    \\
    & & F182M & $7.164 \pm 0.201$ & $8.097 \pm 0.331$ & $4.082 \pm 0.701$
    \\
    & & F200W & $6.248 \pm 0.267$ & $5.718 \pm 0.488$ & $4.632 \pm 0.496$
    \\
    & & F210M & $7.025 \pm 0.245$ & $8.077 \pm 0.416$ & $4.973 \pm 0.712$
    \\
    & & F250M & $5.421 \pm 0.397$ & $6.941 \pm 0.211$ & $3.857 \pm 0.998$
    \\
    & & F277W & $4.624 \pm 0.259$ & $5.101 \pm 0.183$ & $3.309 \pm 0.511$
    \\
    & & F300M & $4.061 \pm 0.231$ & $4.569 \pm 0.166$ & $2.835 \pm 0.729$
    \\
    & & F335M & $5.304 \pm 0.310$ & $5.951 \pm 0.173$ & $3.582 \pm 0.821$
    \\
    & & F356W & $5.008 \pm 0.296$ & $5.514 \pm 0.199$ & $3.787 \pm 0.644$
    \\
    & & F410M & $4.883 \pm 0.513$ & $4.400 \pm 0.320$ & $3.19 \pm 1.10$
    \\
    & & F444W & $3.890 \pm 0.433$ & $3.692 \pm 0.273$ & $1.488 \pm 0.942$
    \\
    & JWST/MIRI & F770W & \multicolumn{3}{c}{$1.60 \pm 2.23$}
    \\
    \cmidrule{2-6}
    UV magnitude $M_\text{UV} \, (\mathrm{mag})$ & & F210M through F444W & $-18.492_{-0.038}^{+0.039}$ & $-18.601_{-0.035}^{+0.036}$ & $-18.03_{-0.13}^{+0.15}$
    \\
    Bolometric luminosity $L_\text{bol} \, (10^{10} \, \mathrm{L_\odot})$ & & F210M through F444W & $9.0_{-1.4}^{+1.6}$ & $9.6_{-1.2}^{+1.4}$ & $7.8_{-3.5}^{+6.6}$
    \\
    UV slope (up to $5 \, \mathrm{\upmu m}$) $\beta_\text{UV}$ & & F210M through F444W & $-2.75_{-0.10}^{+0.10}$ & $-2.719_{-0.082}^{+0.082}$ & $-2.88_{-0.37}^{+0.36}$
    \\
    UV slope (up to $3.5 \, \mathrm{\upmu m}$) $\hat{\beta}_\text{UV}$ & & F210M through F335M & $-3.10_{-0.14}^{+0.14}$ & $-2.81_{-0.13}^{+0.13}$ & $-3.19_{-0.56}^{+0.57}$
    \\
    \bottomrule
\end{tabular}
        \\
        \flushleft
        Reported quantities (and corresponding $1\sigma$ uncertainties) are the flux density $F_\nu$ in nJy, the UV magnitude ($M_\text{UV}$) in magnitudes, the bolometric luminosity ($L_\text{bol}$) in $10^{10}$ Solar luminosity, and UV slopes taking into account all available filters redwards of $2 \, \mathrm{\upmu m}$ ($\beta_\text{UV}$) or only up to and including F335M ($\hat{\beta}_\text{UV}$). Fluxes in available HST and JWST filters are measured with \textsc{forcepho} and within circular $0.3\arcsec$-diameter apertures (CIRC2), except for MIRI/F770W, as detailed in Photometric measurements. Synthetic photometry in NIRCam filters is directly extracted from the NIRSpec/PRISM spectrum (see Supplementary information). For each of the three different sets of photometry, UV properties ($L_\text{bol}$, $M_\text{UV}$ and $\beta_\text{UV}$) are measured redwards of $\lambda_\text{obs} = 2.0 \, \mathrm{\upmu m}$, corresponding to rest-frame wavelengths $\lambda_\text{emit} \gtrsim 1500 \, \Angstrom$ at $z = 13$ (see Supplementary information for details). The uncertainty on the UV magnitude ($M_\text{UV}$) takes into account a systematic uncertainty of $\Delta z = 0.05$.
    \end{table*}
\endgroup

\section*{Methods}

\subsection*{Cosmology and conventions}
\label{ssec:Cosmology_conventions}

A flat $\Lambda$CDM cosmology is adopted throughout based on the latest results of the Planck collaboration\citemet{2020A&A...641A...6P}, with $H_0 = 67.4 \, \mathrm{km \, s^{-1} \, Mpc^{-1}}$, $\Omega_\text{m} = 0.315$, $\Omega_\text{b} = 0.0492$. The cosmic hydrogen fraction is fixed to $f_\text{H} = 0.76$. At $z = 13$, the Hubble flow is $H(z = 13) \approx 1990 \, \mathrm{km \, s^{-1} \, Mpc^{-1}}$, and on-sky separations of $1\arcsec$ and $1\arcmin$ correspond to $3.53 \, \text{physical kpc}$ (pkpc) and $0.212 \, \text{pMpc}$, respectively. We quote magnitudes in the AB system\citemet{1983ApJ...266..713O}, emission-line wavelengths in vacuum, and EWs in the rest frame unless explicitly mentioned otherwise.

\subsection*{NIRCam observations and target selection}
\label{ssec:NIRCam_observations_and_target_selection}

In the following sections, we describe the main JWST and auxiliary Hubble Space Telescope (HST) observations underlying this work. We refer to \citetm{2024ApJ...970...31R} and \citetmet{2024arXiv240518462H} for details on the NIRCam and MIRI imaging, respectively, while \citetm{2024Natur.633..318C} provide a detailed description of the NIRSpec spectroscopy. Further details on the JADES survey strategy and data reduction are discussed in the survey overview paper\citem{2023arXiv230602465E} and the data release papers\citemet{2023ApJS..269...16R, 2024A&A...690A.288B, 2024arXiv240406531D}.

The NIRCam\citem{2023PASP..135b8001R}, MIRI\citem{2015PASP..127..584R} and NIRSpec\citem{2022A&A...661A..80J}$^,$\citemet{2023PASP..135c8001B} measurements presented in this work are associated with JWST guaranteed time observations (GTO) programme IDs (PIDs) 1180 (PI: Eisenstein), 1210, 1286, and 1287 (PI: Luetzgendorf), further complemented with the JOF programme\citem{2023arXiv231012340E} (PID 3215, PIs: Eisenstein \& Maiolino). In addition, since the JOF itself is located within the Great Observatories Origins Deep Survey-South (GOODS-S\citemet{2004ApJ...600L..93G}) extragalactic legacy field, HST Legacy Fields imaging\citemet{2016arXiv160600841I} is publicly available, covering $0.4 \, \mathrm{\upmu m}$ to $1.8 \, \mathrm{\upmu m}$ between the Advanced Camera for Surveys (ACS) and Wide Field Camera 3 (WFC3).

Additional MIRI imaging in the F770W filter was obtained\citemet{2024arXiv240518462H} as coordinated parallel observations to JADES NIRCam observations (PID 1180). Several high-redshift targets, selected by \citetm{2024ApJ...964...71H} and \citetm{2024ApJ...970...31R} based on the NIRCam images in the JOF, including \JGSzthirteenLA\footnote{Located at right ascension of $+53.06475 \, \mathrm{deg}$ and declination of $-27.89024 \, \mathrm{deg}$.}, were followed up using the NIRSpec micro-shutter array (MSA\citemet{2022A&A...661A..81F}) as part of PID 1287, scheduled between 10 and 12 January 2024.

\subsection*{NIRSpec observations and data reduction}
\label{ssec:NIRSpec_observations_and_data_reduction}

The NIRSpec observations spanned three consecutive visits, however during visit 2 the lock on the guide star was lost, preventing it from being carried out nominally. While different MSA configurations were adopted across visits, \JGSzthirteenLA was observed in both visits 1 and 3 in the PRISM/CLEAR grating-filter combination (simply `PRISM' hereafter) with resolving power of $30 \lesssim R \lesssim 300$ between wavelengths of $0.6 \, \mathrm{\upmu m}$ and $5.3 \, \mathrm{\upmu m}$, as well as in the medium-resolution grating-filter combinations G140M/F070LP, G235M/F170LP, and G395M/F290LP (`R1000 gratings'), each with resolving power $R \approx 1000$. A sequence of exposures following three nod positions was repeated four times for each visit in PRISM mode, and once for each of the R1000 gratings. Each nod sequence had an exposure time of $8403.2 \, \mathrm{s}$, consisting of six integrations made up of 19 groups in NRSIRS2 readout mode\citemet{2017PASP..129j5003R}. Altogether, \JGSzthirteenLA was observed for $67225.6 \, \mathrm{s}$ by the NIRSpec/PRISM and $16806.4 \, \mathrm{s}$ in each of the R1000 gratings.

We employed version 3.1 of the data reduction pipeline developed by the ESA NIRSpec Science Operations Team\citemet{2022A&A...661A..81F} and the NIRSpec GTO team (simply `pipeline' hereafter), which produces flux-calibrated spectra largely following the algorithms adopted in the Space Telescope Science Institute (STScI) pipeline\footnote{\url{https://jwst-docs.stsci.edu/jwst-science-calibration-pipeline}}. We refer to previous works\citem{2023NatAs...7..622C, 2024Natur.633..318C}$^,$\citemet{2024A&A...690A.288B, 2024arXiv240406531D} for detailed descriptions of the NIRSpec data reduction pipeline, an overview of which is given in \citetmet{2022A&A...661A..81F}. In brief, three adjacent micro-shutters were opened to obtain background-subtracted spectra of individual sources, where the subtraction follows a three-point nodding scheme discussed above. Initial path-loss corrections were calculated under the assumption of a point-source light profile placed at the same intra-shutter location of the source. The PRISM spectra adopt an irregular wavelength grid with sampling such that the wavelength-dependent LSF\citem{2022A&A...661A..80J} always spans a fixed number of wavelength bins. Our fiducial (`sigma-clipped') spectrum combines all available sub-exposures in the three nodding positions, for which one-dimensional spectra are extracted over the central three spatial pixels (corresponding to $0.3\arcsec$), via a custom sigma-clipping algorithm (see Supplementary information for details).

\subsection*{Photometric measurements}
\label{ssec:Photometric_measurements}

We obtained photometric measurements of \JGSzthirteenLA via two methods. Our fiducial photometry is determined using \textsc{forcepho} (Johnson et~al. in prep.) on all $14$ available NIRCam filters (see also \citetm{2024ApJ...970...31R}), while the MIRI/F770W follows a customised procedure following Helton et~al. (in prep.), both discussed in more detail below. An alternative approach to \textsc{forcepho} is to measure fluxes in circular apertures with a diameter of $0.3\arcsec$ (`CIRC2'). These results are summarised in Extended Data \cref{tab:Photometry}. We include CIRC2 photometry in the available HST bands, which, together with NIRCam filters up to and including F150W, are statistically fully consistent with non-detections ($\chi^2 = 11.6$ over $10$ filters, i.e. $p = 0.31$).

Given the full-width half maximum (FWHM) of the MIRI/F770W point spread function (PSF) is significantly larger than those of NIRCam\citemet{2024arXiv240518462H}, we considered the $\text{F444W}-\text{F770W}$ colour of \JGSzthirteenLA after convolving the F444W mosaic with the F770W PSF and rebinning to the F770W pixel size. We measured this colour assuming a circular aperture with $0.7 \arcsec$ diameter (`CIRC5'), which roughly corresponds to the $65\%$ encircled energy of F770W, prior to applying aperture corrections. The reported MIRI/F770W flux is then inferred from the difference between the total CIRC5 NIRCam/F444W flux and the $\text{F444W}-\text{F770W}$ colour. By adopting this approach, we are taking advantage of the higher spatial resolution afforded by NIRCam compared to MIRI. However, this measurement does not yield a significant detection ($F_\nu = 1.60 \pm 2.23 \, \mathrm{nJy}$). Neglecting contributions from the $\OIII \, \lambda \, 4960, 5008 \, \Angstrom$ lines and underlying continuum, the MIRI non-detection would be consistent with an \Hbeta flux of $F_\text{\Hbeta} \lesssim 6.7 \times 10^{-19} \, \mathrm{erg \, s^{-1} \, cm^{-2}}$ ($3\sigma$), translating to an intrinsic \Lya flux of $F_\text{\Lya} \lesssim 1.6 \times 10^{-17} \, \mathrm{erg \, s^{-1} \, cm^{-2}}$ (case-B recombination; e.g. ref. \citem{2025MNRAS.536...27W}).

To explore the morphology of \JGSzthirteenLA, we first fitted \citetmet{1963BAAA....6...41S} profiles separately to the various available NIRCam filters (using the mosaic images) employing the \textsc{pysersic} code\citemet{2023JOSS....8.5703P}. We do not find a strong wavelength dependency of the morphology. In the F277W filter, which probes rest-frame wavelengths around $\lambda_\text{emit} \approx 2000 \, \Angstrom$ at $z = 13$, we constrain \JGSzthirteenLA to have a half-light radius of $17.5_{-1.7}^{+3.0} \, \mathrm{mas}$ and a \citeauthor{1963BAAA....6...41S} index consistent with $n = 1$. This size approaches half the pixel size (i.e. $15 \, \mathrm{mas}$) and should hence be treated as an upper limit, given the mosaicing procedure likely introduces artificial smoothing.

To fit to independent dithered NIRCam exposures, we performed further modelling with \textsc{forcepho} (Johnson et~al. in prep.), adopting a model with a single intrinsic \citeauthor{1963BAAA....6...41S} profile and freely varying normalisation in each filter (e.g. refs. \citemet{2023NatAs...7..611R, 2023ApJ...952...74T, 2024NatAs.tmp..246B}). Importantly, by fitting to the individual exposures \textsc{forcepho} avoids correlated noise between pixels in drizzled mosaic images, enabling us to probe scales smaller than individual pixels. The results are shown in Extended Data \cref{fig:ForcePho_fit}, and the resulting photometry is listed in Extended Data \cref{tab:Photometry}. From this analysis, we find a formal upper limit (\nth{84} percentile) on the half-light radius of $5.1 \, \mathrm{mas}$. We therefore conclude the imaging data is consistent with the continuum source being unresolved. Based on tests with similarly faint brown dwarf stars that allow the expected systematic uncertainties to be quantified, we conservatively adopt an upper limit on the half-light radius as reported by \citetm{2024ApJ...970...31R} for the F200W filter, $\lesssim 10 \, \mathrm{mas}$ or $\lesssim 35 \, \mathrm{pc}$.
\begin{figure*}
	\centering
	\includegraphics[width=\linewidth]{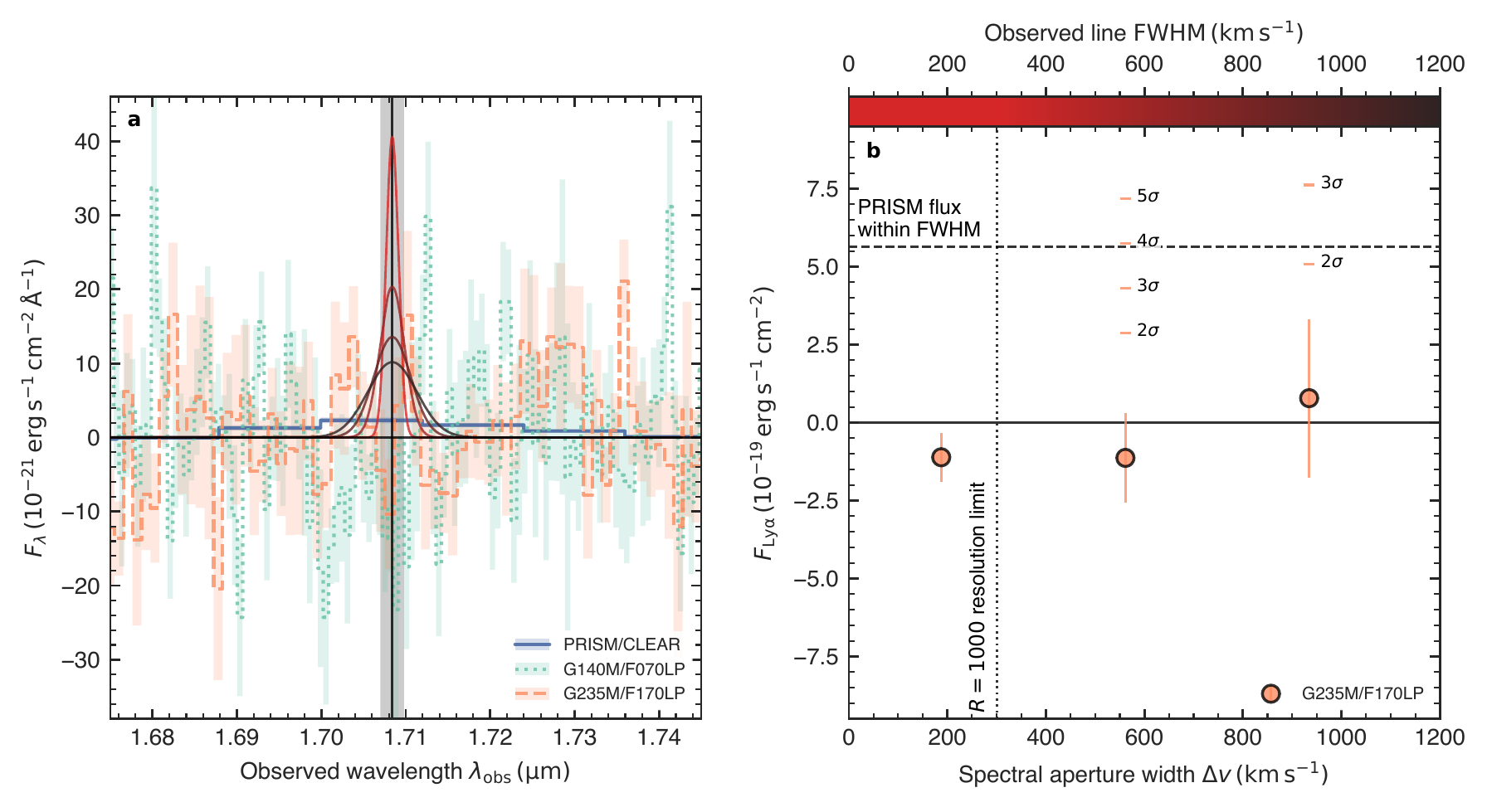}
	\caption{\textbf{Medium-resolution (R1000) grating spectra of \JGSzthirteenLA.} \textbf{a}, Coloured lines represent observed spectra in different grating-filter modes, as obtained from the sigma-clipping procedure (Supplementary information). Specifically, we show the G140M/F070LP (turquoise, dotted) and G235M/F170LP (orange, dashed) spectra compared to the low-resolution PRISM spectrum (dark blue). Shading represents a $1\sigma$ uncertainty on all components of the figure. Solid curves represent emission-line profiles at increasing widths (according to the colourbar in panel~b), starting from the $R = 1000$ resolution limit and having matched the flux and central wavelength ($1.708 \, \mathrm{\upmu m}$; indicated by a vertical black line) to the values measured from the PRISM spectrum (see Emission-line properties). \textbf{b}, Measured \Lya flux in an increasingly wide spectral aperture centred on $1.708 \, \mathrm{\upmu m}$ in G235M/F170LP are shown by circles with $1\sigma$ error bars, none of which show a significant detection. This is consistent with the less sensitive G140M/F070LP measurements (not shown here for clarity). A horizontal dashed line shows the measured PRISM line flux contained within the FWHM of a Gaussian profile (76\%), while a vertical dotted line indicates the limiting $R = 1000$ resolution. This illustrates that if the emission line is well resolved ($\text{FWHM} \gtrsim 600 \, \mathrm{km \, s^{-1}}$), it would fall below the nominal noise level of the R1000 gratings (cf. annotated $2\sigma$ and $4\sigma$ levels).
	}
	\label{fig:Grating_spectra}
\end{figure*}
\begingroup
    \setlength{\tabcolsep}{6pt} 
    \renewcommand{\arraystretch}{1.25} 
    \begin{table}
        \centering
        \caption
        {\textbf{Emission-line constraints for \JGSzthirteenLA.}}
        \label{tab:Emission_lines}
        \begin{tabular}{lcc}
            \toprule
            Emission line(s) & $F \, (10^{-19} \, \mathrm{erg \, s^{-1} \, cm^{-2}})$ & EW ($\Angstrom$)
            \\
            \midrule
            \Lya & $7.42 \pm 1.16$ & $>40^*$
            \\
            $\HeII \, \lambda \, 1640 \, \Angstrom$ & $<1.6$ & $<30$
            \\
            $\NIV \, \lambda \, 1483, 1487 \, \Angstrom$ & $<2.0$ & $<29$
            \\
            $\CIV \, \lambda \, 1548, 1551 \, \Angstrom$ & $<1.6$ & $<27$
            \\
            $\OIII \, \lambda \, 1660, 1666 \, \Angstrom$ & $<1.2$ & $<25$
            \\
            \NIII & $<1.8$ & $<43$
            \\
            \CIIIs & $<1.1$ & $<36$
            \\
            $\OII \, \lambda \, 3727, 3730 \, \Angstrom$ & $<0.95$ & $<42$
            \\
            \bottomrule
        \end{tabular}
        \\
        \flushleft
        Presented quantities for each line are the flux and EW from the PRISM spectra. Constraints for undetected lines are presented as $3\sigma$ upper limits. \NIII refers to the multiplet at $1750 \, \Angstrom$, while \CIIIs is shorthand for $\CIIIf \, \lambda \, 1907 \, \Angstrom, \CIIIs \, \lambda \, 1909 \, \Angstrom$. \\
        $^*$Discussed in more detail in Spectral modelling.
    \end{table}
\endgroup

\subsection*{Emission-line properties}
\label{ssec:Emission-line_properties}

The emission line at $1.71 \, \mathrm{\upmu m}$ is clearly and consistently detected across different PRISM data reductions, even when only one of the two individual visits is considered (Supplementary information). We first fit a Gaussian profile to the sigma-clipped spectrum using the corresponding covariance matrix (Supplementary information), which provides a good fit to the data: $\chi^2 = 5.97$ with $5$ degrees of freedom (DOF). We obtain a centroid of $1.7084 \pm 0.0014 \, \mathrm{\upmu m}$ and $\text{FWHM} = 302 \pm 18 \, \Angstrom$ (or $\Delta v \approx 5000 \, \mathrm{km \, s^{-1}}$) that spans $2.4$ wavelength bins ($120 \, \Angstrom$ wide at $1.71 \, \mathrm{\upmu m}$). We conclude that the line is likely unresolved in the PRISM spectrum and, as expected for compact sources observed with the NIRSpec MSA\citemet{2024A&A...684A..87D}, that the spectral resolution is enhanced by a factor of $\approx 1.5\times$ compared to the resolution curve predicted for a uniformly illuminated micro-shutter.\footnote{Available at \url{https://jwst-docs.stsci.edu/jwst-near-infrared-spectrograph/nirspec-instrumentation/nirspec-dispersers-and-filters}.}

To measure the absolute flux of the line, we first applied a correction to both the sigma-clipped spectrum and the covariance matrix based on the linear \textsc{forcepho} fit found in our path-loss analysis (Supplementary information) to account for additional path losses in the NIRSpec measurements. Directly integrating the corrected PRISM spectrum across the four wavelength bins between $1.69 \, \mathrm{\upmu m}$ and $1.73 \, \mathrm{\upmu m}$ (each bin with $\text{SNR} > 1$; Supplementary Information), we find a flux of $F = 7.42 \pm 1.16 \times 10^{-19} \, \mathrm{erg \, s^{-1} \, cm^{-2}}$ (i.e. the line is detected at $\text{SNR} = 6.4$). We have verified that all different data reductions (see Supplementary information) yield measurements consistent within $1\sigma$. Specifically, the two visits independently confirm the line detection with measured fluxes of $5.77 \pm 1.36 \times 10^{-19} \, \mathrm{erg \, s^{-1} \, cm^{-2}}$ and $9.07 \pm 1.80 \times 10^{-19} \, \mathrm{erg \, s^{-1} \, cm^{-2}}$, respectively.

The emission line is not detected in the medium-resolution G140M/F070LP or G235M/F170LP spectra, both of which cover $1.71 \, \mathrm{\upmu m}$ as shown in Extended Data \cref{fig:Grating_spectra} (though we note the G235M/F170LP transmission drops below $1.7 \, \mathrm{\upmu m}$; ref. \citem{2022A&A...661A..80J}). To quantify whether this is expected, taking into account their inherently lower sensitivity and relatively short exposure times compared to the PRISM (NIRSpec observations and data reduction), we tested if the observed R1000 spectra are consistent with the line flux measured in the PRISM spectra. Indeed, we find that if the observed line profile is sufficiently broadened ($\text{FWHM} \gtrsim 600 \, \mathrm{km \, s^{-1}}$, i.e. well-resolved at $R = 1000$ resolution), it would be below the current sensitivity ($\lesssim 2\sigma$ detection expected; Extended Data \cref{fig:Grating_spectra}).

As discussed further in the Supplementary information, we find it highly unlikely that the emission line at $1.71 \, \mathrm{\upmu m}$ is due to contamination of the micro-shutter by a foreground source that is aligned with \JGSzthirteenLA by chance and remains undetected in the continuum, given that the continuum emission of \JGSzthirteenLA unambiguously places the source at $z \approx 13$. We have performed the `redshift sweep' analysis detailed in the appendices of \citetm{2024ApJ...976..160H} and \citetm{2024Natur.633..318C}, in which the inferred the one-sided $p$-value for a set of different emission lines is combined to yield the statistical significance of a potential spectroscopic confirmation at a given redshift. The effectiveness of this method is illustrated by the case of GS-z14-0, where the most likely redshift was revealed\citem{2024Natur.633..318C} to be $z = 14.178$ (combined $p = 0.0072$) mainly based on a $3.6\sigma$ detection of \CIIIs. This redshift, consistent within the uncertainty determined from fitting the \Lya break profile with DLA absorption, was later independently confirmed via the detection\citem{2024arXiv240920533C}$^,$\citemet{2024arXiv240920549S} of the $\OIII \, 88 \, \mathrm{\upmu m}$ emission line by ALMA. In the case of \JGSzthirteenLA, the redshift sweep was performed across a range of $\Delta z = 0.2$ centred on $z = 13.0$, which however did not show any significant line detections.

Upper limits on the flux and EW for other, undetected lines at $z = 13$ are therefore determined from integrating the covariance matrix across $3$ PRISM wavelength bins, taking into account any residual flux after having subtracted a power-law model continuum (Spectral modelling). The resulting limits, summarised in Extended Data \cref{tab:Emission_lines}, are consistent with findings on most other $z > 10$ galaxies observed by JWST, which generally have revealed these lines to be relatively weak\citem{2023NatAs...7..622C, 2023Natur.622..707A, 2023ApJ...957L..34W, 2024A&A...689A.152D, 2024ApJ...960...56H}.

\begin{figure*}
	\centering
	\includegraphics[width=\linewidth]{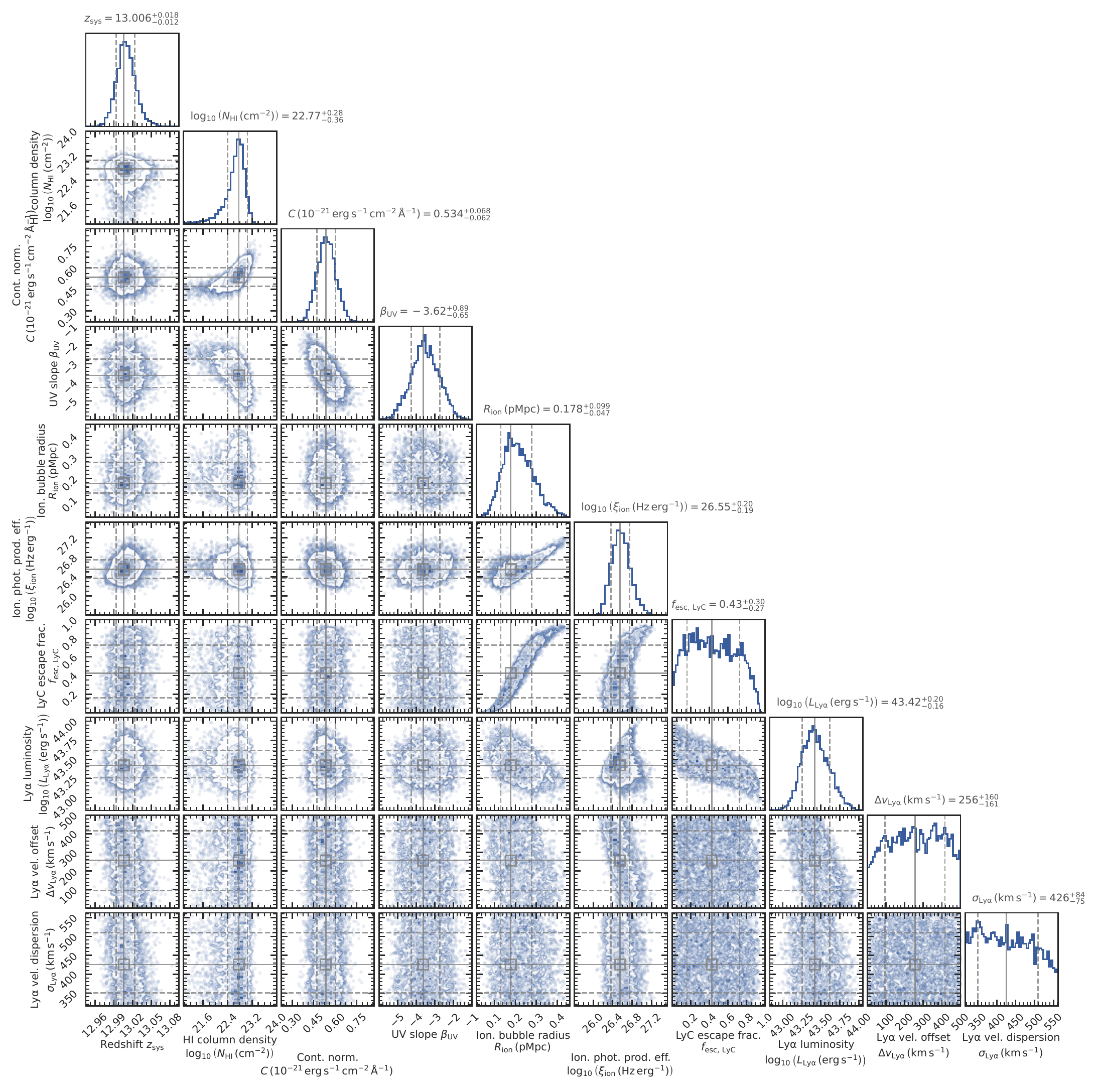}
	\caption{\textbf{Posterior distributions from spectral modelling of the observed spectrum of \JGSzthirteenLA.} The small panels show inter-dependencies between all $8$ parameters freely varied in the model (Extended Data \cref{tab:Model_results}). Additionally, we include the physical radius of the ionised bubble ($R_\text{ion}$) and \Lya luminosity ($L_\text{\Lya}$), which are not independently varied but instead are determined by the other parameters (see Spectral modelling).
	}
	\label{fig:Corner_plot}
\end{figure*}
\begingroup
    \setlength{\tabcolsep}{3.75pt} 
    \renewcommand{\arraystretch}{1.5} 
    \begin{table*}
        \centering
        \caption
        {\textbf{Spectral model parameters, prior distributions, and best-fitting values.}}
        \label{tab:Model_results}
        \begin{tabular}{llllccP{1.65cm}P{1.65cm}P{1.65cm}P{1.65cm}}
            \toprule
            Parameter & (Logarithmic) unit & Type & Prior & Min. & Max. & Default (power law) & Pure $2\gamma$ & Self-consistent & Fixed $R_\text{ion} = 0$
            \\
            \midrule
            $z_\text{sys}$ & & Varied & Uniform & $12.85$ & $13.1$ & $13.01_{-0.01}^{+0.02}$ & $13.01_{-0.02}^{+0.01}$ & $13.01_{-0.02}^{+0.02}$ & $12.99_{-0.02}^{+0.02}$
            \\
            $\log_{10} (N_\text{\HI})$ & $\mathrm{cm^{-2}}$ & Varied & Uniform & $19$ & $24$ & $22.77_{-0.36}^{+0.28}$ & $19.58_{-0.41}^{+1.10}$ & $22.60_{-1.24}^{+0.51}$ & $22.79_{-0.40}^{+0.31}$
            \\
            $C$ & $10^{-21} \, \mathrm{erg \, s^{-1} \, cm^{-2} \, \Angstrom^{-1}}$ & Varied & Uniform & $0$ & $1$ & $0.534_{-0.062}^{+0.068}$ & $0.444_{-0.030}^{+0.031}$ & $0.215_{-0.056}^{+0.104}\,^*$ & $0.546_{-0.081}^{+0.063}$
            \\
            $\beta_\text{UV}$ & & Varied & Uniform & $-6$ & $-1$ & $-3.62_{-0.65}^{+0.89}$ & -- & $-5.52_{-0.43}^{+1.19}$ & $-3.60_{-0.72}^{+0.93}$
            \\
            $R_\text{ion}$ & $\mathrm{pMpc}$ & Coupled/fixed & -- & -- & -- & $0.178_{-0.047}^{+0.099}$ & $0.177_{-0.057}^{+0.102}$ & $0.253_{-0.074}^{+0.092}$ & $0^\dagger$
            \\
            $\log_{10} (\xi_\text{ion})$ & $\mathrm{Hz \, erg^{-1}}$ & Varied & Uniform & $24$ & $28$ & $26.55_{-0.19}^{+0.20}$ & $26.62_{-0.22}^{+0.17}$ & $27.00_{-0.32}^{+0.23}\,^*$ & $26.60_{-0.14}^{+0.09}$
            \\
            $f_\text{esc, LyC}$ & & Varied/fixed & Uniform & $0$ & $1$ & $0.43_{-0.27}^{+0.30}$ & $0.42_{-0.26}^{+0.34}$ & $0.73_{-0.26}^{+0.14}$ & $0^\dagger$
            \\
            $L_\text{\Lya}$ & $10^{43} \, \mathrm{erg \, s^{-1}}$ & Coupled & -- & -- & -- & $2.6_{-0.8}^{+1.5}$ & $2.6_{-0.9}^{+1.5}$ & $1.6_{-0.3}^{+0.3}$ & $5.5_{-0.9}^{+1.9}$
            \\
            $\text{EW}_\text{\Lya, intr}$ & $\Angstrom$ & Coupled & -- & -- & -- & $691_{-254}^{+467}$ & --$^\ddagger$ & $704_{-289}^{+541}$ & $1375_{-420}^{+591}$
            \\
            $\text{EW}_\text{\Lya, obs}$ & $\Angstrom$ & Coupled & -- & -- & -- & $43_{-11}^{+15}$ & --$^\ddagger$ & $67_{-23}^{+43}$ & $41_{-11}^{+16}$
            \\
            $f_\text{esc, \Lya}$ & & Coupled & -- & -- & -- & $0.063_{-0.023}^{+0.026}$ & $0.063_{-0.022}^{+0.027}$ & $0.095_{-0.020}^{+0.025}$ & $0.030_{-0.007}^{+0.008}$
            \\
            $\Delta v_\text{\Lya, int}$ & $\mathrm{km \, s^{-1}}$ & Varied & Uniform & $0$ & $500$ & $256_{-161}^{+160}$ & $257_{-163}^{+161}$ & $379_{-150}^{+83}$ & $391_{-111}^{+74}$
            \\
            $\Delta v_\text{\Lya, obs}$ & $\mathrm{km \, s^{-1}}$ & Coupled & -- & -- & -- & $694_{-299}^{+299}$ & $603_{-269}^{+299}$ & $638_{-299}^{+359}$ & $930_{-299}^{+329}$
            \\
            $\sigma_\text{\Lya}$ & $\mathrm{km \, s^{-1}}$ & Varied & Log-uniform & $10^{2.5}$ & $10^{2.75}$ & $426_{-75}^{+84}$ & $430_{-77}^{+82}$ & $436_{-81}^{+83}$ & $457_{-86}^{+73}$
            \\
            \midrule
            $\chi^2$ & & & & & & $168.1$ & $175.6$ & $171.4$ & $168.4$
            \\
            DOF & & & & & & $119$ & $120$ & $119$ & $120$
            \\
            \bottomrule
        \end{tabular}
        \\
        \flushleft
        Model parameters are the systemic redshift ($z_\text{sys}$), DLA neutral hydrogen column density ($N_\text{\HI}$), power-law continuum normalisation ($C$) and slope ($\beta_\text{UV}$), ionising-photon production efficiency ($\xi_\text{ion}$) and escape fraction ($f_\text{esc, LyC}$), and the peak velocity offset ($\Delta v_\text{\Lya, int}$) and intrinsic velocity dispersion ($\sigma_\text{\Lya}$) of \Lya line profile emerging from the galaxy. Additional reported parameters are the ionised bubble radius ($R_\text{ion}$), \Lya luminosity ($L_\text{\Lya}$), \Lya EW as it emerges from the galaxy ($\text{EW}_\text{\Lya, intr}$) and as it is observed (i.e. after IGM transmission; $\text{EW}_\text{\Lya, obs}$), \Lya escape fraction ($f_\text{esc, \Lya}$), and the observed \Lya velocity offset ($\Delta v_\text{\Lya, obs}$), which are not freely varied but derived from the main parameters. Best-fitting values (uncertainties) are the median (\nth{16} and \nth{84} percentiles) of the posterior distribution under the default (power-law) model, a model with pure two-photon continuum ($2\gamma$), a self-consistent model incorporating power-law and nebular-emission components, and a power-law model where $R_\text{ion} = 0$ (for details, see Spectral modelling).
        \\
        $^*$ (Relative to the) power-law continuum only.
        \\
        $^\dagger$ Value is fixed in this model.
        \\
        $^\ddagger$ The $2\gamma$ continuum tends to zero approaching the wavelength of \Lya.
    \end{table*}
\endgroup

\subsection*{Spectral modelling}
\label{ssec:Spectral_modelling}

To gain insight into the \Lya emission and absorption properties of \JGSzthirteenLA, we model the observed spectrum with a simple framework in which \Lya and continuum emission produced inside the central galaxy are subject to (damping-wing) absorption arising in intervening neutral hydrogen in dense absorbing systems and/or the IGM. We emphasise that the aim of this model is not to be as physically detailed as possible, which would involve performing simulations including three-dimensional radiative transfer coupled to the hydrodynamics of the gas (requiring the relevant feedback processes to be accurately modelled), but rather to constrain the basic physical properties that \JGSzthirteenLA must possess to explain the observations.

As we expect the \Lya line to be redshifted with respect to the systemic redshift of the galaxy (potentially already as \Lya emerges from the galaxy or otherwise resulting from processing by the neutral IGM\citemet{2024MNRAS.531.2701T, 2024ApJ...972...56T}) and no other emission lines are detected (Emission-line properties), this quantity ($z_\text{sys}$) is not precisely known and is a free parameter in this model. To remain agnostic about the nature of the ionising source and to avoid the intrinsic limitations of standard SPS models in reproducing very blue UV continua (Supplementary information), the continuum emission is modelled as a power law, $F_\lambda \propto \lambda^{\beta_\text{UV}}$, by default. This introduces two more free parameters in the model, the UV slope $\beta_\text{UV}$ and a normalisation (at rest-frame wavelength of $\lambda_\text{emit} = 1500 \, \Angstrom$).

To reproduce the smooth \Lya break seen in the continuum, we allow the continuum emission to be impacted by DLA absorption parametrised by the neutral hydrogen column density $N_\text{\HI}$ as in refs. \citem{2024A&A...689A.152D, 2024ApJ...976..160H}. The \Lya emission is explicitly not attenuated by this absorption, as this would completely extinguish the line.\footnote{Since the attenuated continuum tends to zero at the wavelength of \Lya, we calculate the line EW according to the unattenuated continuum, which is effectively equivalent to measuring the continuum level via the photometry.} As discussed in the main text, this would require a specific geometrical configuration such that the \Lya emission is not strongly absorbed, however \Lya emission superimposed on DLA troughs has been observed in galaxy spectra, suggesting that these geometries exist\citem{2023ApJ...956...39H, 2024MNRAS.532.4703W}. The absorption cross section of neutral hydrogen is based on the Voigt profile approximation given by \citetmet{2006ApJ...645..792T}, with a quantum-mechanical correction provided by \citetmet{2015MNRAS.446..264B}. Since we find that the redshift of the foreground DLA system (when freely varied; e.g. ref. \citemet{2024A&A...690A..70T}) prefers a solution close to the systemic redshift, $z_\text{DLA} \approx z_\text{sys}$, for simplicity we fix $z_\text{DLA} = z_\text{sys}$ in the following.

Alternatively, we considered the case where the observed spectrum is dominated by the $2\gamma$ continuum, which has a fixed shape\citemet{1951ApJ...114..407S} and thus only requires one free parameter, the normalisation. As a third variant, we considered a combination consisting of a power-law continuum (using the same parametrisation as above) and a full nebular emission spectrum, which in addition to the $2\gamma$ continuum and the \Lya line also contains the free-bound (and free-free) components. The nebular emission in this case was computed with the \textsc{pyneb} code\citemet{2015A&A...573A..42L}, which however requires assuming the gas temperature and density. We opted for $T = \num{20000} \, \mathrm{K}$ and $n = 100 \, \mathrm{cm^{-3}}$ respectively, where the $2\gamma$ continuum is dominant contributor the wavelength range considered here\citemet{2016PASP..128k4001S}. The choice for this relatively low density is motivated by the fact that the free-bound (and free-free) components mainly contribute at longer wavelengths and would have to be subdominant to reproduce the very steep UV slope. In this multi-component (`self-consistent') model, we tied the continuum normalisation to the strength of the \Lya line, thereby self-consistently scaling the continuum according to the production rate and escape fraction of LyC photons discussed below.

Following refs. \citem{2025MNRAS.536...27W}$^,$\citemet{2024A&A...682A..40W}, IGM transmission was calculated with the patchy reionisation model presented in \citetmet{2020MNRAS.499.1395M}, integrating along the trajectory of a photon which starts in an ionised bubble of radius $R_\text{ion}$ located in an otherwise neutral IGM (see also refs. \citemet{2000ApJ...542L..75C, 2004ApJ...613...23M}). Following \citetmet{2020MNRAS.499.1395M}, we assume the gas in the ionised bubble to be highly ionised (residual neutral fraction fixed at $x_\text{\HI} = 10^{-8}$) and have $T = 10^4 \, \mathrm{K}$, whereas the neutral IGM is at $T = 1 \, \mathrm{K}$. The gas in both media is assumed to be at mean cosmic density (i.e. to have $\bar{n}_\text{H} \approx 5.25 \times 10^{-4} \, \mathrm{cm^{-3}}$ at $z = 13$), and be at rest with respect to the central source. We fixed the global neutral hydrogen fraction of the IGM (i.e. outside the ionised bubble\citemet{2024A&A...682A..40W}) to $\bar{x}_\text{\HI} = 1$, motivated by various types of evidence which consistently indicate that globally, the Universe is still highly neutral well below redshift $z = 13$ (e.g. refs. \citemet{2019MNRAS.485.3947M, 2024ApJ...975..208T}).

\begin{figure}
	\centering
	\includegraphics[width=\linewidth]{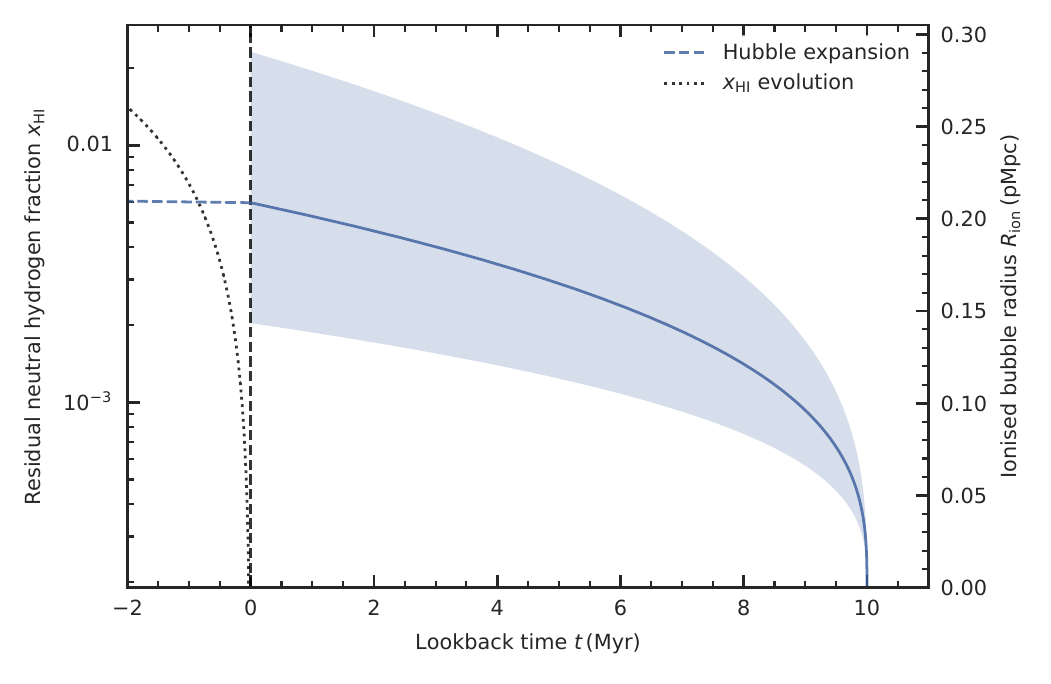}
	\caption{\textbf{Modelled ionised bubble size evolution.} The right axis shows the physical radius of the ionised bubble $R_\text{ion}$, whose evolution as a function of lookback time $t$ is governed by \cref{eq:dR_iondt}. The solid line shows the median among the posterior distribution of the default model, shading represents $1\sigma$ uncertainty (\nth{16} to \nth{84} percentile). The dashed line illustrates the Hubble expansion rate if the bubble remains unchanged from $t = 0$ onwards, showing this effect has little impact over the timescale relevant to our analysis. The dotted line shows how the neutral hydrogen fraction within the bubble (left axis) would evolve without additional ionising photons.
	}
	\label{fig:R_ion_evolution}
\end{figure}

We self-consistently model the size of the ionised bubble by considering the production rate and escape fraction of hydrogen-ionising photons of the central galaxy. As in \citetm{2025MNRAS.536...27W}, we define $\xi_\text{ion} \equiv \dot{N}_\text{ion} / L_\mathrm{\nu, \, UV}$, where $\dot{N}_\text{ion}$ is the production rate of ionising photons and $L_\mathrm{\nu, \, UV}$ is the luminosity density (in units of $\mathrm{erg \, s^{-1} \, Hz^{-1}}$) of the intrinsic continuum of the ionising source at $\lambda_\text{emit} = 1500 \, \Angstrom$. In the case of the multi-component model in particular, $L_\mathrm{\nu, \, UV}$ is taken to be the value of the power-law continuum at $1500 \, \Angstrom$ such that $\xi_\text{ion}$ reflects the intrinsic value. The rate of ionising photons leaking from the galaxy at a given production efficiency $\xi_\text{ion}$ is modulated by the LyC escape fraction, $f_\text{esc, LyC}$. In a given model instance, we therefore begin by deriving the rate of ionising photons escaping the galaxy via (e.g. refs. \citem{2015ApJ...802L..19R}$^,$\citemet{2013ApJ...768...71R, 2019ApJ...879...36F, 2020ApJ...892..109N})
\begin{equation}
    \dot{N}_\text{ion, esc} = f_\text{esc, LyC} \, \dot{N}_\text{ion} = f_\text{esc, LyC} \, \xi_\text{ion} \, L_\mathrm{\nu, \, UV} \, .
\end{equation}
\noindent To calculate the bubble radius $R_\text{ion}$, we then numerically integrate equation (3) in \citealt{2000ApJ...542L..75C}, describing the time evolution of $R_\text{ion} (t)$ to obey
\begin{equation}
    \label{eq:dR_iondt}
    \frac{d R_\text{ion}^3}{dt} = 3 H(z) R_\text{ion}^3 + \frac{3 \dot{N}_\text{ion, esc}}{4\pi \bar{n}_\text{H}} - C_\text{\HII} \bar{n}_\text{H} \alpha_\text{B} R_\text{ion}^3 \, ,
\end{equation}
thereby taking into account the effect of the expansion of the Universe parametrised by the Hubble parameter $H(z)$ and recombinations within the ionised bubble, for which we assume a clumping factor for ionised gas of $C_\text{\HII} = 3$ (e.g. ref. \citemet{2009MNRAS.394.1812P}) and case-B recombination rate $\alpha_\text{B}$ at $\num{20000} \, \mathrm{K}$ as given by \citetmet{2011piim.book.....D}. The typical recombination timescale at $z = 13$, $\left( C_\text{\HII} \bar{n}_\text{H} \alpha_\text{B} \right)^{-1} \approx 140 \, \mathrm{Myr}$, indicates that \JGSzthirteenLA as an ionising source could quickly ionise its surroundings before recombinations are able to restore balance. As illustrated in Extended Data \cref{fig:R_ion_evolution}, showing the time evolution of $R_\text{ion}$ in the default model, the bubble radius can reach $R_\text{ion} \approx 0.1 \, \mathrm{pMpc}$ over a timescale of only $1 \, \mathrm{Myr}$. We note that when the supply of LyC photons ceases, the residual neutral hydrogen fraction does rapidly increase due to the high density at $z = 13$ ($x_\text{\HI} \approx 0.01$ after $1 \, \mathrm{Myr}$), implying that for an ionised bubble to have a significant transmission-enhancing effect redwards of the systemic \Lya wavelength\citemet{2020MNRAS.499.1395M} it must be actively maintained. Here, we integrate until reaching a fiducial age of $t = 10 \, \mathrm{Myr}$, having verified that changing this assumption has little impact on our findings as a result of the sub-linear scaling $R_\text{ion} \propto t^{1/3}$ (in the absence of recombinations and the Hubble flow). We additionally considered an alternative model identical to the default power-law one, but where we fix $R_\text{ion} = 0$ (i.e. $f_\text{esc, LyC} = 0$).

Finally, we determine the intrinsic \Lya luminosity resulting from recombinations by considering the number of ionising photons that are absorbed within the galaxy and reprocessed into \Lya. Similarly to the above, the effective rate of LyC photons contributing to the recombination rate within the galaxy ($\dot{N}_\text{rec}$) follows from the product of the intrinsic production rate $\dot{N}_\text{ion}$ and absorbed fraction (one minus the escape fraction) of ionising photons. This is multiplied by the fraction of (case-B) recombination events that result in the emission of a \Lya photon, $f_\text{rec, B}$ (see e.g. ref. \citemet{2021A&A...650A..98W}), to arrive at the emission rate of \Lya photons, and hence the \Lya luminosity (i.e. the emission rate times the energy of a \Lya photon),
\begin{align}
    \label{eq:Lya_luminosity}
    L_\text{\Lya} & = \dot{N}_\text{rec} \, f_\text{rec, B} \, E_\text{\Lya} \nonumber
    \\ & = \left( 1 - f_\text{esc, LyC} \right) \xi_\text{ion} \, L_\mathrm{\nu, \, UV} \, f_\text{rec, B} \, E_\text{\Lya} \, .
\end{align}

We used $f_\text{rec, B} (T = \num{20000} \, \mathrm{K}) = 0.647$ based on the prescription by \citetmet{2014PASA...31...40D}, noting it depends only weakly on temperature\citemet{2006agna.book.....O} and that case A would lead to an unaccounted increase in $f_\text{esc, LyC}$. Under the very conservative assumptions of no IGM absorption at all and $f_\text{esc, LyC} = 0$, \cref{eq:Lya_luminosity} places a lower limit on the LyC production efficiency via the observed \Lya luminosity relative to the continuum, yielding $\xi_\text{ion} \gtrsim 10^{25.1} \, \mathrm{Hz \, erg^{-1}}$ ($10^{25.4} \, \mathrm{Hz \, erg^{-1}}$ under case A). The \Lya line, as it emerges from the galaxy, is modelled as a Gaussian profile with a given velocity dispersion $\sigma_\text{\Lya}$, which is shifted in velocity space at a given offset from the systemic redshift, $\Delta v_\text{\Lya, int}$, and normalised to the \Lya luminosity derived as described above.

Radiative transfer calculations predict a large variety of \Lya spectral profiles may emerge from galaxies\citemet{2016ApJ...826...14G, 2023MNRAS.523.3749B}, but galactic outflows typically cause systematically redshifted components\citemet{2006A&A...460..397V, 2008A&A...491...89V}, as seen ubiquitously at high redshift\citemet{2018ApJ...856....2M, 2021MNRAS.503.4105T, 2021MNRAS.508.1686W, 2021MNRAS.505.1382M, 2024MNRAS.531.2701T, 2024ApJ...972...56T}. While the emergent \Lya spectral profile is fundamentally unknown at $z \gtrsim 7$ due to the asymmetric IGM transmission on the blue side\citemet{2020MNRAS.499.1395M, 2024A&A...682A..40W}, some clues are given by the non-detection of the line in the R1000 spectra. If the line were unresolved at a resolution of $R \approx 1000$, that is $\text{FWHM} \lesssim 300 \, \mathrm{km \, s^{-1}}$, we would have likely seen a marginal detection (Extended Data \cref{fig:Grating_spectra}). Instead, the line profile likely contains a prominent red, broad component to allow for sufficient transmission of \Lya flux at $z = 13$ even in the presence of an ionised bubble (\cref{fig:NIRSpec_model}). We note that due to the IGM transmission, the peak of the intrinsic line profile at velocity offset $\Delta v_\text{\Lya, int}$ with respect to systemic effectively gets further redshifted to a velocity offset of $\Delta v_\text{\Lya, obs}$.

We employed the \textsc{pymultinest}\citemet{2014A&A...564A.125B} implementation of the multimodal nested-sampling algorithm \textsc{multinest}\citemet{2009MNRAS.398.1601F} to perform a Bayesian fitting routine\footnote{Code available at \url{https://github.com/joriswitstok/lymana_absorption}.} to the sigma-clipped PRISM spectrum and corresponding covariance matrix (see Supplementary information) from $1.609 \, \mathrm{\upmu m}$ up to $2.897 \, \mathrm{\upmu m}$ ($127$ wavelength bins), or $1150 \, \Angstrom \lesssim \lambda_\text{emit} \lesssim 2000 \, \Angstrom$ at $z = 13$. Before fitting, as in Emission-line properties we corrected the NIRSpec measurements for additional path losses. Meanwhile, the model spectrum is convolved with the PRISM resolution curve predicted for a uniformly illuminated micro-shutter, enhanced by a factor of $1.5$ based on the measured width of the \Lya line in the PRISM (see Emission-line properties). As detailed in Jakobsen et~al. (in prep.), the goodness of fit statistic $\chi^2$ is calculated as the matrix product
\begin{equation}
    \chi^2 = \mathbf{R}^T \mathbf{\Sigma}^{-1} \mathbf{R} \, ,
\end{equation}
where $\mathbf{\Sigma}^{-1}$ is the inverted covariance matrix and the $i^\text{th}$ element of the vector $\mathbf{R}$ is given as the difference between observed flux density in the $i^\text{th}$ wavelength bin ($F_{\lambda, i}^\text{obs}$) and the modelled one ($F_{\lambda, i}^\text{model}$),
\begin{equation}
    R_i = F_{\lambda, i}^\text{obs} - F_{\lambda, i}^\text{model} \, .
\end{equation}

The model log-likelihood $\ell$ is calculated assuming the observed data are normally distributed around the model, $\ell = -\frac{1}{2} \chi^2$. All model parameters, adopted prior distributions, and resulting best-fitting values are summarised in Extended Data \cref{tab:Model_results}. The posterior distributions for the default model are shown in Extended Data \cref{fig:Corner_plot}.

While the multi-component self-consistent model has a slightly higher $\chi^2$ ($171.4$) than the default power-law model ($\chi^2 = 168.1$), interestingly it favours a high LyC escape fraction ($f_\text{esc, LyC} = 0.73_{-0.26}^{+0.14}$) to suppress the nebular continuum, much like the SPS model fits (Supplementary information). Indeed, fixing $R_\text{ion} = 0$ in the self-consistent model (results not included here) yields a significantly poorer fit ($\chi^2 = 183.1$), as this overpredicts the continuum tied to the strong \Lya line. Moreover, the intrinsic \Lya flux required for the $R_\text{ion} = 0$ power-law model is discrepant at a $4.5\sigma$ level with the MIRI/F770W non-detection (Photometric measurements).

\section*{Data availability}

The NIRCam data that support the findings of this study are publicly available at \url{https://archive.stsci.edu/hlsp/jades}. The reduced spectra that support the findings of this study are publicly available at \url{https://doi.org/10.5281/zenodo.14714293}.

\section*{Code availability}

The code used for the Spectral modelling fitting routine is available at \url{https://github.com/joriswitstok/lymana_absorption}. The \textsc{astropy}\citemet{2013A&A...558A..33A, 2018AJ....156..123A} software suite is publicly available, as is \textsc{bagpipes}\citemet{2018MNRAS.480.4379C}, \textsc{beagle}\citemet{2016MNRAS.462.1415C}, \textsc{cloudy}\citemet{2017RMxAA..53..385F}, \textsc{emcee}\citemet{2013PASP..125..306F}, \textsc{forcepho}\citemet{2024ascl.soft10006B}, \textsc{multinest}\citemet{2009MNRAS.398.1601F}, \textsc{pymultinest}\citemet{2014A&A...564A.125B}, \textsc{pyneb}\citemet{2015A&A...573A..42L}, the \textsc{SciPy} library\citemet{Jones2001}, its packages \textsc{NumPy}\citemet{2011CSE....13b..22V} and \textsc{Matplotlib}\citemet{Hunter2007}, and \textsc{SpectRes}\citemet{2017arXiv170505165C}.


\bibliographymet{GS-z13-1-LA}

\section*{Acknowledgements}

We thank Laura Keating, Harley Katz, Callum Witten, William McClymont, Arjen van der Wel, John Chisholm, Danielle Berg, and Masami Ouchi for useful discussions. This work is based on observations made with the National Aeronautics and Space Administration (NASA)/European Space Agency (ESA)/Canadian Space Agency (CSA) JWST. The data were obtained from the Mikulski Archive for Space Telescopes at the STScI, which is operated by the Association of Universities for Research in Astronomy, Inc., under NASA contract NAS 5-03127 for JWST. These observations are associated with programmes 1180, 1210, 1286, 1287, and 3215. JW, RM, WMB, FDE, and JS acknowledge support from the Science and Technology Facilities Council (STFC), by the European Research Council (ERC) through Advanced Grant 695671 ``QUENCH'', by the UK Research and Innovation (UKRI) Frontier Research grant RISEandFALL. JW also gratefully acknowledges support from the Cosmic Dawn Center through the DAWN Fellowship. The Cosmic Dawn Center (DAWN) is funded by the Danish National Research Foundation under grant No. 140. BDJ, BER, FS, PAC, DJE, CNAW, and YZ acknowledge support from the JWST/NIRCam contract to the University of Arizona, NAS5-02015. BER also acknowledges support from JWST Program 3215. ST acknowledges support by the Royal Society Research Grant G125142. AJC, AJB, AS, JC, and GCJ acknowledge funding from the ``FirstGalaxies'' Advanced Grant from the ERC under the European Union’s Horizon 2020 research and innovation programme (Grant agreement No. 789056). RS acknowledges support from a STFC Ernest Rutherford Fellowship (ST/S004831/1). SAl acknowledges support from the JWST MIRI Science Team Lead, grant 80NSSC18K0555, from NASA Goddard Space Flight Center to the University of Arizona. SAr acknowledges grant PID2021-127718NB-I00 funded by the Spanish Ministry of Science and Innovation/State Agency of Research (MICIN/AEI/10.13039/501100011033). This research is supported in part by the Australian Research Council Centre of Excellence for All Sky Astrophysics in 3 Dimensions (ASTRO 3D), through project number CE170100013. SCa acknowledges support by European Union’s HE ERC Starting Grant No. 101040227 ``WINGS''. ECL acknowledges support of an STFC Webb Fellowship (ST/W001438/1). DJE is supported as a Simons Investigator. PGP-G acknowledges support from grant PID2022-139567NB-I00 funded by Spanish Ministerio de Ciencia e Innovaci\'on MCIN/AEI/10.13039/501100011033, FEDER, UE. H{\"U} acknowledges funding by the European Union (ERC APEX, 101164796). Views and opinions expressed are however those of the authors only and do not necessarily reflect those of the European Union or the European Research Council Executive Agency. Neither the European Union nor the granting authority can be held responsible for them. The research of CCW is supported by NOIRLab, which is managed by the Association of Universities for Research in Astronomy (AURA) under a cooperative agreement with the National Science Foundation. This study made use of the Prospero high performance computing facility at Liverpool John Moores University. This version of the article has been accepted for publication, after peer review (when applicable) but is not the Version of Record and does not reflect post-acceptance improvements, or any corrections. The Version of Record is available online at: \url{https://dx.doi.org/10.1038/s41586-025-08779-5}.

\newpage

\section*{Author information}

\subsection*{Affiliations}
\noindent
\hypertarget{inst:Kavli}$^{1}$Kavli Institute for Cosmology, University of Cambridge, Madingley Road, Cambridge CB3 0HA, UK
\\
\hypertarget{inst:Cav}$^{2}$Cavendish Laboratory, University of Cambridge, 19 JJ Thomson Avenue, Cambridge CB3 0HE, UK
\\
\hypertarget{inst:DAWN}$^{3}$Cosmic Dawn Center (DAWN), Copenhagen, Denmark
\\
\hypertarget{inst:NBI}$^{4}$Niels Bohr Institute, University of Copenhagen, Jagtvej 128, DK-2200, Copenhagen, Denmark
\\
\hypertarget{inst:UCL}$^{5}$Department of Physics and Astronomy, University College London, Gower Street, London WC1E 6BT, UK
\\
\hypertarget{inst:Steward}$^{6}$Steward Observatory, University of Arizona, 933 N. Cherry Avenue, Tucson AZ 85721, USA
\\
\hypertarget{inst:CfA}$^{7}$Center for Astrophysics $|$ Harvard \& Smithsonian, 60 Garden St., Cambridge MA 02138, USA
\\
\hypertarget{inst:UCSC}$^{8}$Department of Astronomy and Astrophysics University of California, Santa Cruz, 1156 High Street, Santa Cruz CA 96054, USA
\\
\hypertarget{inst:Oxford}$^{9}$Department of Physics, University of Oxford, Denys Wilkinson Building, Keble Road, Oxford OX1 3RH, UK
\\
\hypertarget{inst:LJMU}$^{10}$Astrophysics Research Institute, Liverpool John Moores University, 146 Brownlow Hill, Liverpool L3 5RF, UK
\\
\hypertarget{inst:CAB}$^{11}$Centro de Astrobiolog\'ia (CAB), CSIC–INTA, Cra. de Ajalvir Km.~4, 28850- Torrej\'on de Ardoz, Madrid, Spain
\\
\hypertarget{inst:ESAC}$^{12}$European Space Agency (ESA), European Space Astronomy Centre (ESAC), Camino Bajo del Castillo s/n, 28692 Villanueva de la Cañada, Madrid, Spain
\\
\hypertarget{inst:SNS}$^{13}$Scuola Normale Superiore, Piazza dei Cavalieri 7, I-56126 Pisa, Italy
\\
\hypertarget{inst:IAP}$^{14}$Sorbonne Universit\'e, CNRS, UMR 7095, Institut d'Astrophysique de Paris, 98 bis bd Arago, 75014 Paris, France
\\
\hypertarget{inst:ESO}$^{15}$European Southern Observatory, Karl-Schwarzschild-Strasse 2, 85748 Garching, Germany
\\
\hypertarget{inst:Herts}$^{16}$Centre for Astrophysics Research, Department of Physics, Astronomy and Mathematics, University of Hertfordshire, Hatfield AL10 9AB, UK
\\
\hypertarget{inst:INAF}$^{17}$INAF -- Osservatorio Astronomico di Brera, via Brera 28, I-20121 Milano, Italy
\\
\hypertarget{inst:AURA}$^{18}$AURA for European Space Agency, Space Telescope Science Institute, 3700 San Martin Drive. Baltimore, MD 21210, USA
\\
\hypertarget{inst:Wisconsin}$^{19}$Department of Astronomy, University of Wisconsin-Madison, 475 N. Charter St., Madison WI 53706, USA
\\
\hypertarget{inst:MPE}$^{20}$Max-Planck-Institut für extraterrestrische Physik, Gießenbachstraße 1, 85748 Garching, Germany
\\
\hypertarget{inst:NOIRLab}$^{21}$NSF's National Optical-Infrared Astronomy Research Laboratory, 950 North Cherry Avenue, Tucson AZ 85719, USA
\\
\hypertarget{inst:NRC}$^{22}$NRC Herzberg, 5071 West Saanich Rd, Victoria BC V9E 2E7, Canada

\subsection*{Author contributions}

JW and PJ led the analysis and the writing of this paper, with key contributions from AJB, AJC, AS, BDJ, BER, FS, JMH, MC, RM, RS, SCa, and ST. AJB, CW, FDE, GCJ, JC, JW, KB, MC, NK, PJ, RM, SAr, SCa, and SCh contributed to the development and commissioning of the NIRSpec instrument and the reduction and analysis of the NIRSpec data presented. BDJ, BER, CCW, CNAW, DJE, FS, KNH, PAC and ST contributed to the development and commissioning of the NIRCam instrument and the reduction and analysis of the NIRCam data presented. JMH and SAl contributed to the reduction and analysis of the MIRI data presented. AJB, BDJ, BER, CCW, CNAW, CW, DJE, ECL, FDE, H{\"U}, JC, JS, KNH, MVM, PGPG, PJ, and RM contributed to the design and execution of the JADES program. AJB, BER, CW, DJE, and ST serve on the JADES Steering Committee. RB, WMB, PR, and YZ provided comments on the manuscript.

\subsection*{Correspondence}

Correspondence should be addressed to J. Witstok.

\section*{Ethics declarations}

\subsection*{Competing interests}

The authors declare no competing interests.

\clearpage
\makeatletter
\renewcommand{\fnum@figure}{Supplementary Material Fig. \thefigure}
\renewcommand{\fnum@table}{Supplementary Material Table \thetable}
\makeatother
\setcounter{page}{1}
\setcounter{figure}{0}
\setcounter{table}{0}
\setcounter{footnote}{0}

\section*{Supplementary information}

\subsection*{NIRCam imaging}
\label{ssec:NIRCam_imaging}

The NIRCam imaging data set in the JOF comprises $14$-band imaging of uniform depth, each having $5\sigma$ limits below $\sim 3 \, \mathrm{nJy}$ for a point source\citem{2023arXiv231012340E, 2024ApJ...970...31R}. A false-colour image centred on \JGSzthirteenLA created from stacked NIRCam image mosaics is shown in Supplementary Material \cref{fig:NIRCam_imaging}. The blue channel is the mean of the F090W and F115W filters, green similarly combines F150W and F162M, while red is a stack of F182M, F200W, and F210M (panel~d of \cref{fig:NIRCam_NIRSpec} was created similarly with all available filters). Photometric redshifts of nearby galaxies have been derived with \textsc{eazy}\citesup{2008ApJ...686.1503B} as outlined in \citetm{2024ApJ...964...71H}.
\begin{figure*}
	\centering
	\includegraphics[width=\linewidth]{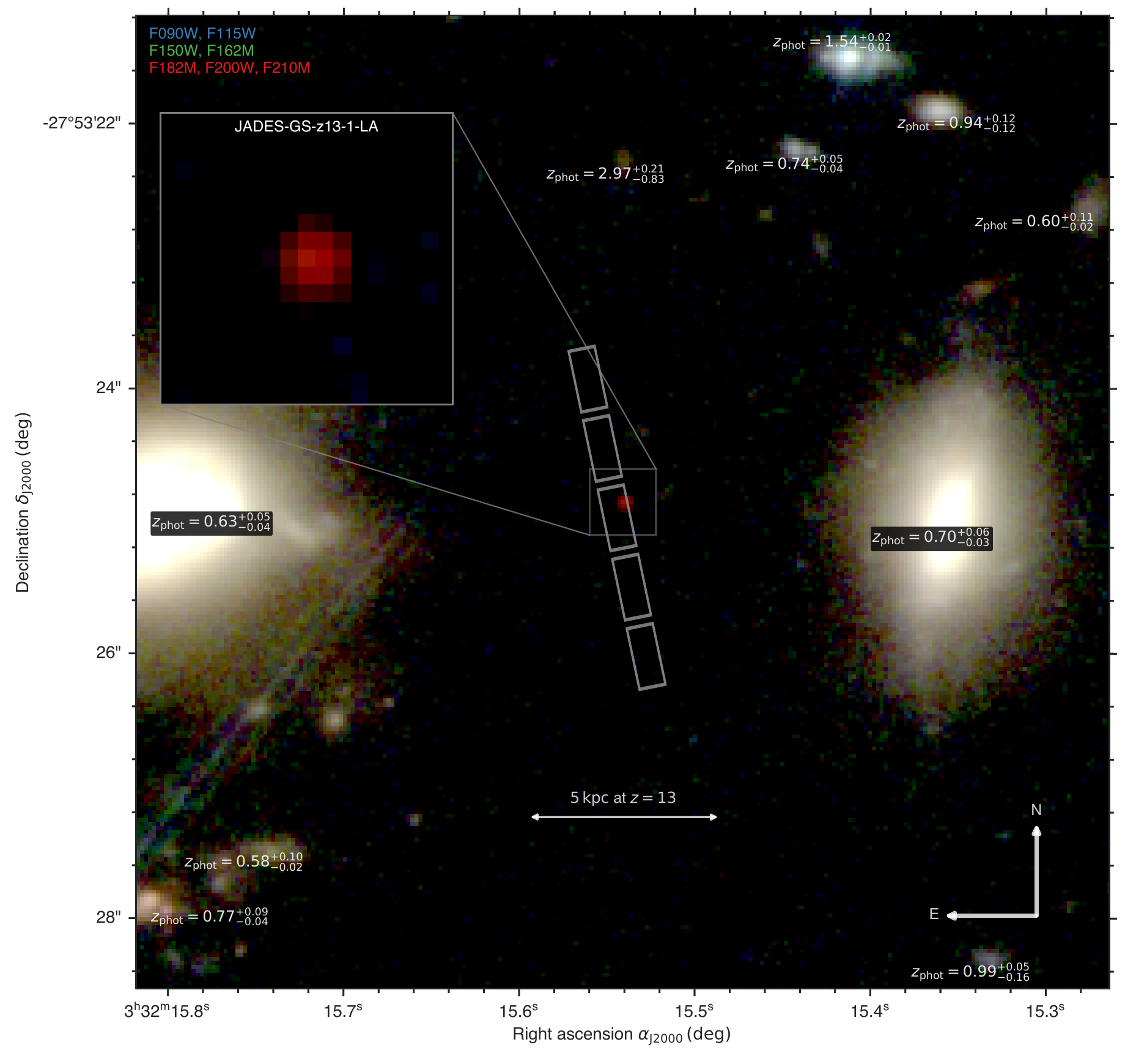}
	\caption{\textbf{NIRCam false-colour image of \JGSzthirteenLA and its surroundings.} The image shows stacks of the F090W and F115W filters as the blue channel, the F150W and F162M filters as green, and F182M, F200W, and F210M as red. To the bottom left, a diffraction spike can be seen stretching diagonally across the image, coming from a star located just outside the field of view. Annotations show the photometric redshifts of nearby sources, including two extended low-redshift galaxies. The placement of the NIRSpec micro-shutters is shown in grey. At the bottom centre, a scale of $5 \, \mathrm{kpc}$ at $z = 13.0$ is indicated.
	}
	\label{fig:NIRCam_imaging}
\end{figure*}

\subsection*{NIRSpec data reduction}
\label{ssec:NIRSpec_data_reduction}

We considered various reduction variations, including one that reduces self-subtraction in cases where the emission is extended over multiple micro-shutters by only considering the two outer nod positions instead of the default three. We have verified that the differences between this reduction and the default one considering all nod positions are minimal, including for the emission line observed at $\lambda_\text{obs} \approx 1.7 \, \mathrm{\upmu m}$. The same is true for reduction variants where one-dimensional spectra are extracted over the central three or five spatial pixels (corresponding to $0.3\arcsec$ or $0.5\arcsec$ respectively), as expected given the compactness of \JGSzthirteenLA (Methods). We therefore base our analysis on 3-pixel extractions considering all three nodding positions to maximise the SNR. Additional path losses are accounted for through a comparison with NIRCam photometry, as discussed in the Methods.
\begin{figure*}
	\centering
	\includegraphics[width=0.95\linewidth]{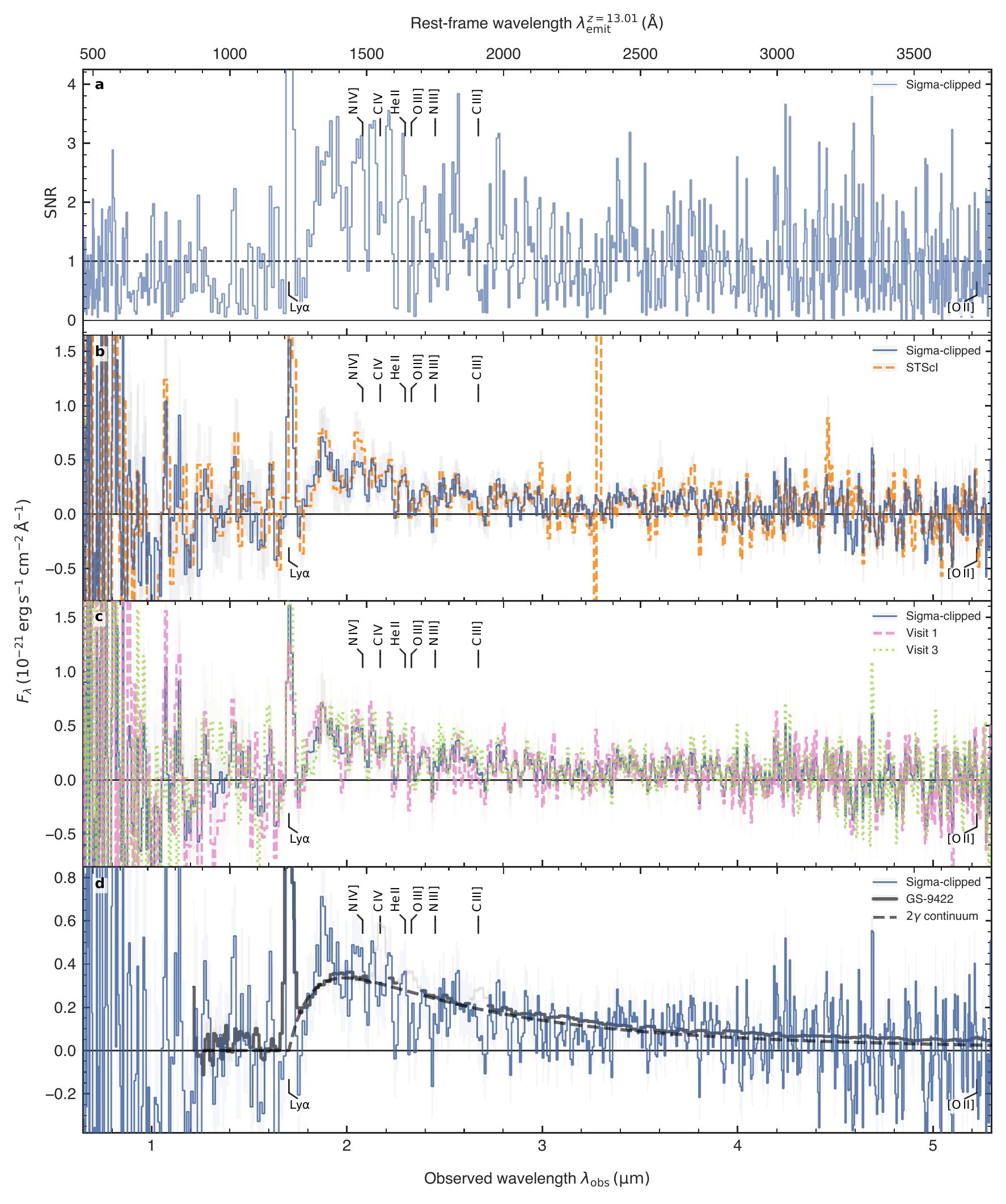}
	\caption{\textbf{Comparison of NIRSpec/PRISM spectra of \JGSzthirteenLA.} \textbf{a}, SNR on individual wavelength bins for the sigma-clipped spectrum combining all sub-exposures (also displayed in other panels). The uncertainty on individual spectral bins considered here only represents diagonal elements of the covariance matrix. \textbf{b}, Comparison with the default STScI pipeline reduction (see NIRSpec observations and data reduction for details). \textbf{c}, Comparison with spectra of the two visits separately. \textbf{d}, Comparison of the spectrum of \JGSzthirteenLA with that of GS-9422 (strong UV lines greyed out to ease comparison) and the two-photon ($2\gamma$) continuum (both rescaled; see text for details). Shading in panels b through d shows $1\sigma$ uncertainty. All panels show the expected locations of key UV emission lines: \Lya (detected), \NIV, \CIV, \HeII, \OIIIs, \NIII, \CIIIs, and \OII (all undetected).
	}
	\label{fig:Comparison_spectra}
\end{figure*}
\begin{figure}
	\centering
	\includegraphics[width=\linewidth]{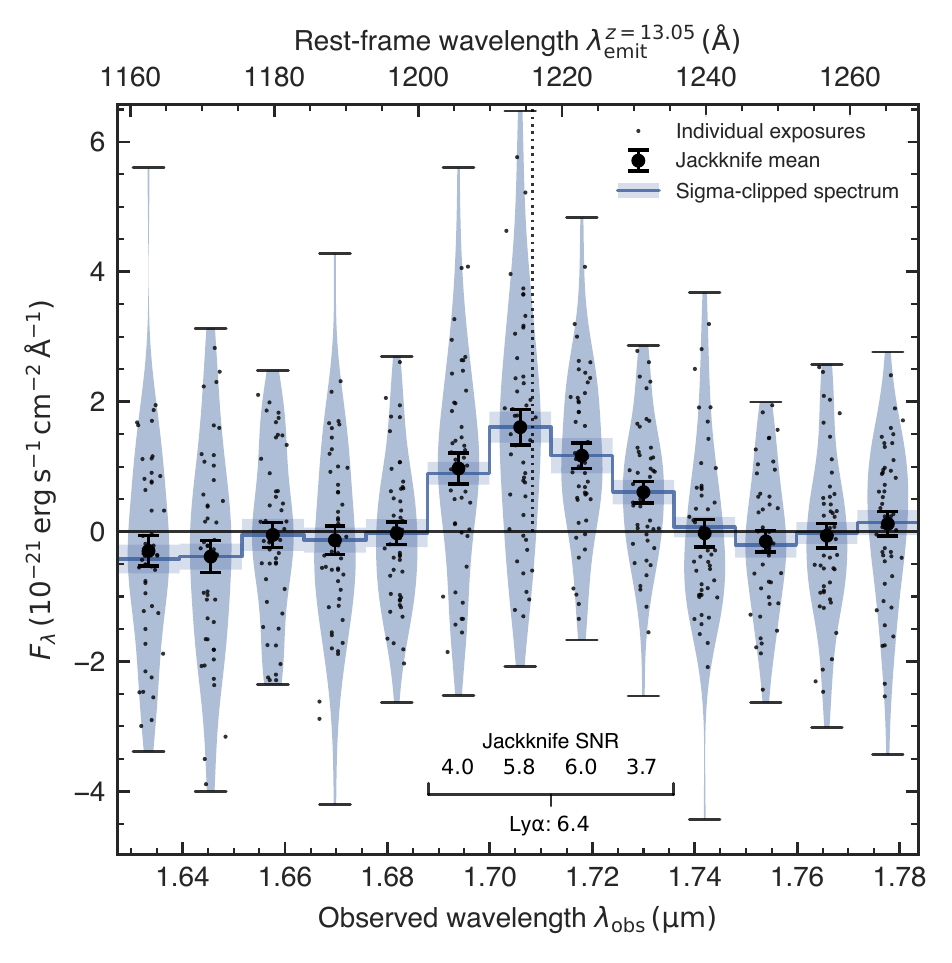}
	\caption{\textbf{Individual sub-spectra of \JGSzthirteenLA around $1.7 \, \mathrm{\upmu m}$.} Our fiducial, sigma-clipped spectrum is shown by the step-wise blue line, with shading representing the $1\sigma$ uncertainty for each individual spectral bin as retrieved from the diagonal element of the covariance matrix. The flux values extracted from individual sub-exposures are shown as small black circles within each wavelength bin (slightly offset in the wavelength direction for visualisation). Their distribution is illustrated with a kernel-density estimate shown by a blue swath, enclosed by short horizontal black bars indicating the flux extrema. Larger black circles represent their mean, with errorbars showing the $68\%$ confidence interval estimated via jackknife resampling. The jackknife SNR is annotated for four individual bins with $\text{SNR} > 1$, as well as for the combined flux across these four bins.
	}
	\label{fig:Individual_exposures_Lya}
\end{figure}

The pipeline independently produces two- and one-dimensional spectra from all reduced sub-exposures, of which there are a substantial number for \JGSzthirteenLA ($48$ for the PRISM and $12$ for each of the R1000 gratings). Since cosmic ray impacts and noisy pixels are not always fully captured in the ramp fitting\citemet{2022A&A...661A..81F} and occasionally introduce a sharp excess of flux in these individual sub-exposures, the standard reduction pipeline adopts iterative sigma clipping\citemet{2024A&A...690A.288B} for each wavelength bin before co-adding the reduced one-dimensional spectra. This combination process weights individual sub-spectra inversely by the pipeline error estimate squared. The combination of two-dimensional spectra is performed separately\citemet{2024A&A...690A.288B}, and therefore the final combined two-dimensional spectrum (as shown in \cref{fig:NIRCam_NIRSpec}) is not used to directly extract one-dimensional spectra. Recently, however, several refinements to this standard procedure were presented in a detailed investigation\citem{2024ApJ...976..160H} (see also ref. \citesup{2024arXiv240702575C}) of ultra-deep NIRSpec/PRISM measurements of two $z > 10$ galaxies benefiting from a large number of sub-exposures ($186$ and $138$ respectively), which we follow here to obtain our fiducial (`sigma-clipped') spectrum.

Specifically, \citetm{2024ApJ...976..160H} introduced two additional masking steps aimed at removing obviously spurious flux values at the beginning of the combination process. In each step, the median flux value and median pipeline error estimate for each wavelength bin are calculated across available sub-exposures. Wavelength bins of sub-spectra where flux values deviate by more than five times the median error from the median flux, or where the pipeline error estimate exceeds five times its median, were eliminated. Subsequently, five iterations of sigma clipping removed flux values deviating from the average of the surviving entries in the bin by more than three times the sample standard deviation. Finally, the individual sub-spectra were co-added simply by averaging the surviving entries in each wavelength bin. As in \citetm{2024ApJ...976..160H} and \citetsup{2024arXiv240702575C}, we constructed $2000$ bootstrapped realisations of the final combined spectrum from which we derived a covariance matrix. Crucially, this covariance matrix directly captures all possible sources of statistical fluctuations present in the data (including those not explicitly taken into account by the default pipeline) as well as the significant degree of correlation in the noise of adjacent wavelength bins.

We note that while the intra-shutter location of \JGSzthirteenLA is nearly identical across the two visits (\cref{fig:NIRCam_NIRSpec}), the sub-exposures follow a three-point nodding pattern, and the employed micro-shutters differ by one column between the two visits (quadrant 3, row 237, column 87 versus quadrant 3, row 237, column 86). As a result, the two-dimensional spectra, including the emission line at $\lambda_\text{obs} \approx 1.7 \, \mathrm{\upmu m}$, are shifted vertically by $\sim 15$ pixels on the detector over the range of the $48$ sub-exposures, thus minimising the potential impact of bad pixels. We have performed careful visual inspection of the raw NIRSpec exposure frames to verify the spectrum does not suffer from overlapping targets or contamination from other (failed) open micro-shutters. Finally, there are no signs of hot pixels or significant particle hits affecting the line at $\lambda_\text{obs} \approx 1.7 \, \mathrm{\upmu m}$ in the extracted spectrum: in the four wavelength bins where the line appears in the combined PRISM spectrum, the refined sigma-clipping algorithm described above has classified $46$ out of $48$ values as valid for one bin, while all 48 values in the remaining three bins were deemed valid.

\subsection*{NIRSpec/PRISM spectra}
\label{ssec:PRISM_spectra}

In Supplementary Material \cref{fig:Comparison_spectra}, we compare different one-dimensional PRISM spectra of \JGSzthirteenLA, including our fiducial spectrum, obtained from the refined sigma-clipping procedure applied on 3-pixel extractions of all available sub-exposures (NIRSpec data reduction), and the default STScI pipeline spectrum. From this comparison, we conclude these generally agree very well, although a number of noise spikes can be seen in the STScI spectrum. We also show spectra obtained separately from the two visits, which both show good agreement with the combined spectrum, including for the emission line observed at $\lambda_\text{obs} \approx 1.7 \, \mathrm{\upmu m}$.

Furthermore, we provide a comparison with the NIRSpec/PRISM spectrum of GS-9422 at $z = 5.94$ (JADES-GS+53.12175-27.79763 in \citetmet{2024A&A...690A.288B}; see also refs. \citem{2024arXiv240402194T}$^,$\citemet{2024A&A...690A..70T}$^,$\citesup{2024ApJ...969L...5L}), whose spectrum simultaneously displays strong \Lya emission, a Balmer jump, and a steep turnover seen in the UV (rest-frame wavelengths below $\lambda_\text{emit} \lesssim 1500 \, \Angstrom$). To guide the eye, Supplementary Material \cref{fig:Comparison_spectra} shows the $2\gamma$ continuum for gas at electron density of $n_e = 10^3 \, \mathrm{cm^{-3}}$ and electron temperature of $T_e = \num{20000} \, \mathrm{K}$ (though we note that unlike the normalisation, the shape of the $2\gamma$ continuum does not depend on these conditions\citemet{1951ApJ...114..407S}), which, like the spectrum of GS-9422, is shifted in wavelength to $z = 13$ and rescaled to match the observed flux density of \JGSzthirteenLA at $\lambda_\text{emit} = 1500 \, \Angstrom$.

A detailed view of the PRISM wavelength bins near the emission line observed at $\lambda_\text{obs} \approx 1.7 \, \mathrm{\upmu m}$ is shown in Supplementary Material \cref{fig:Individual_exposures_Lya}. We performed jackknife resampling of all sub-spectra to independently estimate the uncertainty on the mean flux measured in each wavelength bin across all sub-exposures (in the absence of any masking). These are in excellent agreement the fiducial, sigma-clipped spectrum and uncertainty from the bootstrapped covariance matrix (NIRSpec data reduction), suggesting the measurements are regularly distributed without extreme outliers. We also computed the (unmasked) integrated flux across four bins with $\text{SNR} > 1$ for each sub-spectrum separately. The resulting $\text{SNR} = 6.4$ on the integrated flux estimated using jackknife resampling again agrees well with the SNR estimated via the sigma-clipped spectrum and corresponding bootstrapped covariance matrix, as will be discussed in Emission-line properties. The STScI reduction, on the other hand, nominally returns a higher significance for the line detection ($\text{SNR} = 8.5$), likely reflecting a slight underestimation of the (correlated) noise (NIRSpec data reduction).

We consider here whether the emission line at $1.71 \, \mathrm{\upmu m}$ may be due to contamination of the micro-shutter by a foreground source that is aligned with \JGSzthirteenLA by chance and remains undetected in the continuum. Firstly, such a low-redshift interloper must have an exceptionally high EW to respect the stringent non-detections in the bluest NIRCam filters (e.g. the F150W $3\sigma$ constraint implies an observed $\text{EW} \gtrsim \num{10000} \, \Angstrom$ if the continuum were flat in $F_\nu$). Therefore, the line could realistically only be \Halpha at $z \approx 1.60$, where \Paalpha from the same foreground source would be observed at $4.88 \, \mathrm{\upmu m}$, or $\OIII \, \lambda \, 5008 \, \Angstrom$ at $z \approx 2.41$ where the \Halpha and \NII complex would fall at $2.24 \, \mathrm{\upmu m}$. The absence of any accompanying line detections in the first case implies $\Halpha/\Paalpha > 13.8$ ($2\sigma$), in tension with the case-B ratio\citemet{2006agna.book.....O} of $\Halpha/\Paalpha \approx 10$ which can only be decreased by dust attenuation. Alternatively, for the interloper to be at $z \approx 2.41$ requires a line ratio ($\OIII/(\Halpha+\NII) > 7.7$ at $2\sigma$) completely incompatible with observed properties of $z \sim 2$ galaxies\citesup{2015ApJ...801...88S}. Leaving aside EW arguments, for the line to be the $\OII \, \lambda \, 3727, 3730 \, \Angstrom$ doublet (\OII) at $z \approx 3.58$, similarly unprecedented line ratios would be required ($\OII/\Hbeta > 8.4$ and $\OIII/\OII < 0.11$ at $2\sigma$).

We conclude the only consistent explanation is that the line is \Lya at $z \approx 13$. However, we note its observed properties also cannot be due to the extended \Lya photon-diffusion emission predicted to surround galaxies before reionisation: it has been shown this should extend over a physical size of $\approx 1 \, \mathrm{pMpc}$\citem{1999ApJ...524..527L} and as a result, the diffuse \Lya emission has extremely low surface brightness\citesup{2024JCAP...10..059P}, orders of magnitude lower than observed in the spectrum of \JGSzthirteenLA taken over a solid angle of $0.199\arcsec \times 0.461\arcsec$ spanned by a micro-shutter on sky\citem{2022A&A...661A..80J}. Moreover, our nodding background subtraction technique would largely self-subtract this signal that is predicted to have approximately uniform central surface brightness\citem{1999ApJ...524..527L}.
\begin{figure*}
	\centering
	\includegraphics[width=\linewidth]{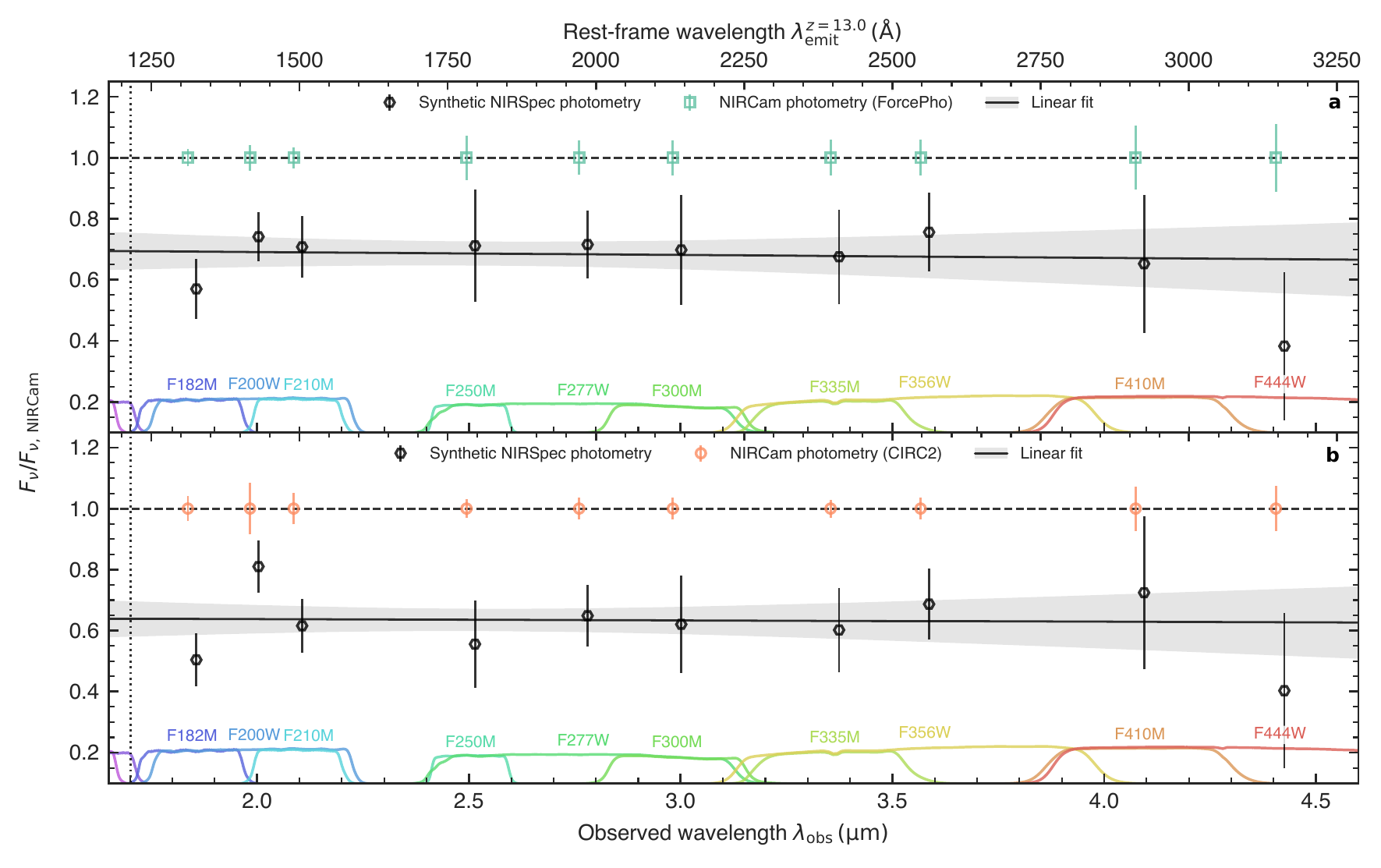}
	\caption{\textbf{Comparison between NIRCam and NIRSpec (spectro)photometry of \JGSzthirteenLA.} Synthetic NIRSpec photometry (see Synthetic photometry and path-loss corrections) is plotted relative to the \textsc{forcepho} (panel \textbf{a}) and CIRC2 (panel \textbf{b}) photometry in each available NIRCam filter redwards of the observed wavelength of \Lya at $z = 13$ (dotted vertical line).
	}
	\label{fig:Path-loss_corrections}
\end{figure*}

\subsection*{Synthetic photometry and path-loss corrections}
\label{ssec:Synthetic_photometry_and_path-loss_corrections}

To assess the quality of the NIRSpec spectrophotometry and quantify any additional path losses not already accounted for by the data reduction pipeline (see NIRSpec observations and data reduction), we created `synthetic' NIRSpec photometry to directly compare against the available NIRCam measurements (cf. Extended Data \cref{tab:Photometry}). Noting all variations of the NIRSpec data reduction yield consistent results, including between the 5-pixel and 3-pixel extractions (Supplementary information), we calculated synthetic flux densities by convolving the sigma-clipped spectrum (with only the default point-source path-loss correction applied; see NIRSpec observations and data reduction) with the NIRCam filter profiles. The corresponding uncertainties were derived using the full covariance matrix and therefore take into account correlated noise properties directly estimated from the measurements. A comparison between the different photometry variants considered here, as derived from NIRSpec (synthetic) and NIRCam (\textsc{forcepho} and CIRC2; see Photometric measurements), is provided in Supplementary Material \cref{fig:Path-loss_corrections}.

We used the \textsc{emcee}\citemet{2013PASP..125..306F} package to perform a linear fit to the ratios of the synthetic NIRSpec photometry to the NIRCam photometry (both the \textsc{forcepho} and CIRC2 measurements) as a function of wavelength. Between \textsc{forcepho} and CIRC2, we consistently find the NIRSpec fluxes are approximately $0.7\times$ those measured by NIRCam (a difference of about $0.4 \, \mathrm{mag}$). There is no clear dependence on the observed wavelength: in the \textsc{forcepho} case, the linear fit has a slope of $-0.01 \pm 0.06 \, \mathrm{\upmu m^{-1}}$, and the CIRC2 photometry has $0.00 \pm 0.05 \, \mathrm{\upmu m^{-1}}$. We note that the point-source approximation used by the pipeline to apply the initial path-loss corrections (NIRSpec observations and data reduction) should be valid in the case of \JGSzthirteenLA, as it is essentially unresolved by NIRCam (see Photometric measurements). However, these corrections are subject to a systematic uncertainty on the intra-shutter position introduced by the limited pointing accuracy of the MSA, which is anticipated\citem{2022A&A...661A..80J} to be of the order of $25 \, \mathrm{mas}$ (i.e. comparable to or larger than the size of \JGSzthirteenLA). This effect is enhanced for sources close to the micro-shutter edge (as is the case for \JGSzthirteenLA; \cref{fig:NIRCam_NIRSpec}), where the path-loss correction shows the largest gradient\citemet{2022A&A...661A..81F}.

\subsection*{UV magnitude and slope}
\label{ssec:UV_magnitude_and_slope}

We first measured the absolute UV magnitude $M_\text{UV}$ and slope $\beta_\text{UV}$ directly from the different sets of photometry. To perform an unbiased inference of these parameters, we consider filters redwards of $\lambda_\text{obs} = 2.0 \, \mathrm{\upmu m}$ (i.e. from F210M on), corresponding to rest-frame wavelengths $\lambda_\text{emit} \gtrsim 1500 \, \Angstrom$ at $z = 13$, thus avoiding the strong spectral break and line (\cref{fig:NIRCam_NIRSpec}). The results are tabulated Extended Data \cref{tab:Photometry}. Given the uncertainty of the systemic redshift (as discussed in the Spectral modelling section of the Methods), reported uncertainties on the UV magnitude conservatively take into account a systematic uncertainty of $\Delta z = 0.05$. The UV slope is reported for fits taking into account all available NIRCam filters, covering the rest frame at $\lambda_\text{emit} \lesssim 3500 \, \Angstrom$. Alternatively, since empirical measurements of UV slopes in the literature use slightly different wavelength ranges (e.g. refs. \citem{2022ApJ...941..153T}$^,$\citesup{2023MNRAS.520...14C, 2024MNRAS.529.4087T, 2024MNRAS.531..997C, 2024arXiv240410751A}), we also report the UV slope only using NIRCam filters below $\lambda_\text{obs} \approx 3.5 \, \mathrm{\upmu m}$ (i.e. up to and including F335M), corresponding to $\lambda_\text{emit} \lesssim 2500 \, \Angstrom$. This reflects the commonly adopted prescription by \citetsup{1994ApJ...429..582C} and more closely mimics the fitting range discussed in Spectral modelling.

The UV magnitude inferred from the (uncorrected) NIRSpec measurements is $0.4$-$0.5 \, \mathrm{mag}$ fainter than those from NIRCam, reflecting the discrepancy found in Synthetic photometry and path-loss corrections. Based on the CIRC2 photometry, we find a UV magnitude of $M_\text{UV} = -18.66_{-0.04}^{+0.04}$, in good agreement with $M_\text{UV} = -18.73 \pm 0.04$ reported by \citetm{2024ApJ...970...31R} (corrected for differences in the adopted cosmology) based on \citetsup{1980ApJS...43..305K} photometry. Among the three sets of photometry considered here, our best-fitting UV slopes are consistently very steep ($\beta_\text{UV} \lesssim -2.7$), again agreeing with $\beta_\text{UV} = -2.73 \pm 0.13$ found by \citetm{2024ApJ...970...31R} when considering the full wavelength range observed by NIRCam ($\lambda_\text{emit} \lesssim 3500 \, \Angstrom$). We find a considerably steeper slope yet ($\beta_\text{UV} < -3$) in the wavelength range up to $\lambda_\text{emit} \lesssim 2500 \, \Angstrom$ (though with larger uncertainties, as a result of the smaller wavelength range and number of data points). Finally, we provide an estimate of the bolometric luminosity ($L_\text{bol}$) by integrating the power law with a lower bound at $\lambda_\text{\HeII} = 227.84 \, \Angstrom$ (i.e. the \HeII ionisation edge at $54.4 \, \mathrm{eV}$, where stellar SEDs are typically suppressed\citem{2022ARA&A..60..455E}).
\begingroup
    \setlength{\tabcolsep}{6pt} 
    \renewcommand{\arraystretch}{1.25} 
    \begin{table}
        \centering
        \caption
        {\textbf{Stellar properties of \JGSzthirteenLA.}}
        \label{tab:SED_models}
        \begin{tabular}{lccc}
            \toprule
            \multirow{2}{*}{Quantity} & \multicolumn{2}{c}{\textsc{bagpipes}} & \textsc{beagle}
            \tabularnewline
            & Default & No nebular & $f_\text{esc, LyC} \neq 0$
            \tabularnewline
            \midrule
            $M_* \, (10^{7} \, \mathrm{M_\odot})$ & $5.5_{-1.2}^{+2.1}$ & $2.7_{-0.7}^{+1.8}$ & $2.6_{-1.5}^{+3.5}$
            \tabularnewline
            $\Sigma_* \, (10^{3} \, \mathrm{M_\odot \, pc^{-2}})$ & $>7.1_{-1.6}^{+2.8}$ & $>3.5_{-0.9}^{+2.3}$ & $>3.3_{-2.0}^{+4.6}$
            \tabularnewline
            $Z_* \, (\mathrm{Z_\odot})$ & $0.3_{-0.1}^{+0.8}\%$ & $0.6_{-0.3}^{+1.1}\%$ & $2_{-1}^{+10}\%$
            \tabularnewline
            $\mathrm{SFR}_{10} \, (\mathrm{M_\odot \, yr^{-1}})$ & $0.19_{-0.17}^{+0.51}$ & $1.84_{-0.62}^{+0.55}$ & $0.67_{-0.18}^{+0.36}$
            \tabularnewline
            $\Sigma_\text{SFR} \, (\mathrm{M_\odot \, yr^{-1} \, kpc^{-2}})$ & $>25_{-21}^{+66}$ & $>240_{-81}^{+71}$ & $>87_{-23}^{+47}$
            \tabularnewline
            $t_* \, (\mathrm{Myr})$ & $22_{-6}^{+9}$ & $11_{-6}^{+13}$ & $21_{-15}^{+38}$
            \tabularnewline
            $A_V \, (\mathrm{mag})$ & $0.04_{-0.03}^{+0.04}$ & $0.11_{-0.07}^{+0.08}$ & $0.08_{-0.05}^{+0.19}$
            \tabularnewline
            $\log_{10} U$ & $-1.91_{-0.79}^{+0.82}$ & -- & $-2.59_{-0.86}^{+0.91}$
            \tabularnewline
            $f_\text{esc, LyC}$ & $0^*$ & $1^*$ & $0.81_{-0.32}^{+0.14}$
            \tabularnewline
            \bottomrule
        \end{tabular}
        \\
        \flushleft
        Best-fitting parameters, taken as the median of their marginalised posterior, are shown for the \textsc{bagpipes} and \textsc{beagle} models with different assumptions on nebular emission (see Stellar population synthesis modelling). Error bars represent a $1 \sigma$ uncertainty (\nth{16} and \nth{84} percentiles). Rows: stellar mass ($M_*$) in $10^7$ Solar masses, stellar mass surface density ($\Sigma_*$) in $10^3$ Solar masses per square parsec, stellar metallicity ($Z_*$) in units of Solar metallicity, star formation rate in Solar masses per year averaged on a timescale of $10 \, \mathrm{Myr}$ ($\text{SFR}_{10}$), its corresponding surface density ($\Sigma_\text{SFR}$) in Solar masses per year per square kiloparsec, mass-weighted stellar age ($t_*$) in $\mathrm{Myr}$, visual dust extinction ($A_V$) in magnitudes, ionisation parameter $U$ (if applicable), and ionising-photon escape fraction ($f_\text{esc, LyC}$).
        \\
        $^*$ Value is fixed in this model.
    \end{table}
\endgroup
\begin{figure*}
	\centering
	\includegraphics[width=\linewidth]{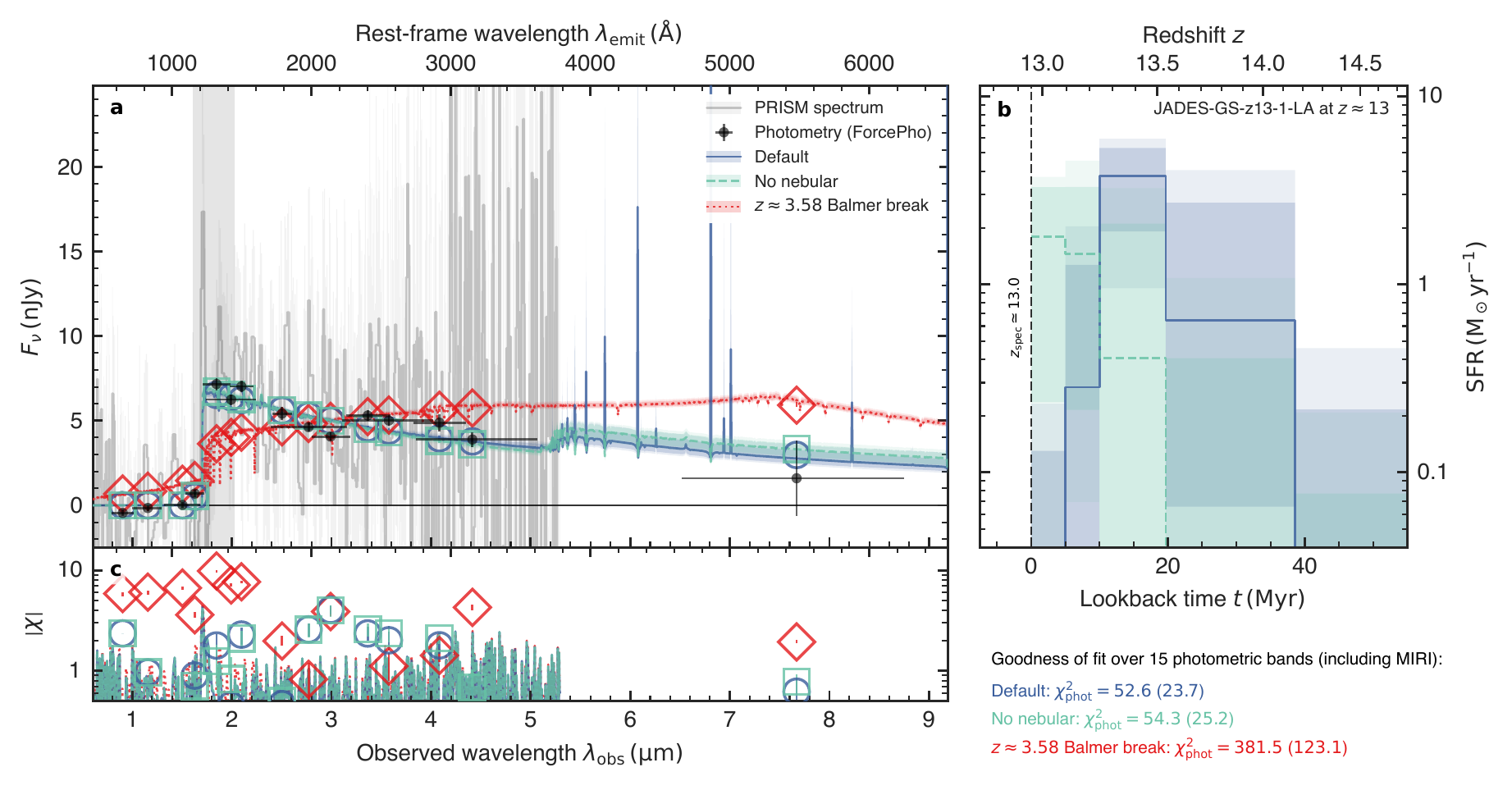}
	\caption{\textbf{SED modelling of \JGSzthirteenLA.} Three different SPS models produced by \textsc{bagpipes} are shown: the default setup (circles), a model without nebular emission (`No nebular', squares), and a forced low-redshift model ($z \approx 3.58$ Balmer break, diamonds; see Stellar population synthesis modelling for details). \textbf{a}, Observed NIRCam (\textsc{forcepho}) and MIRI photometry (black points) and NIRSpec/PRISM spectrum (grey line; smoothed by a $3$-pixel median filter for visualisation), overlaid by the three models (lines according to the legend in the top right and open symbols). Error bars on the photometry and light-grey shading around the spectrum represent $1\sigma$ uncertainties; darker (lighter) shading around the model lines shows $1\sigma$ ($2\sigma$) variation in the posterior distribution. \textbf{b}, Residuals between the observations and modelled spectra and photometry expressed in terms of the absolute value of $\chi$, the difference normalised by $1\sigma$ observational uncertainty. Goodness-of-fit statistics in the form of $\chi^2$ calculated over the $15$ photometric bands (including the MIRI/F770W measurement) are annotated (conservative estimates in brackets). \textbf{c}, Inferred SFH in the two $z \approx 13$ models (default and without nebular emission; shading as in panel~a).
	}
	\label{fig:SED_models}
\end{figure*}

\subsection*{Stellar population synthesis modelling}
\label{ssec:SPS_modelling}

We modelled the full SED of \JGSzthirteenLA taking into account NIRCam photometry\footnote{We have verified that including the MIRI/F770W photometric measurement does not noticeably change our findings, so we chose not to include it in the fitting results shown here unless explicitly mentioned.} as well as the NIRSpec/PRISM measurements (the 3-pixel extraction from the default pipeline) with standard SPS models. Specifically, we follow the procedure outlined in \citetm{2025MNRAS.536...27W}, using the Bayesian Analysis of Galaxies for Physical Inference and Parameter EStimation (\textsc{bagpipes}\citemet{2018MNRAS.480.4379C}) set up with v2.2.1 Binary Population and Spectral Synthesis (\textsc{bpass}\citesup{2017PASA...34...58E}) binary-star models, assuming the default \textsc{bpass} IMF (stellar mass upper limit of $300 \, \mathrm{M_\odot}$). We masked the NIRSpec measurements in the spectral region around \Lya, $1150 \, \Angstrom < \lambda_\text{emit} < 1450 \, \Angstrom$, where the complex spectral shape featuring a strong \Lya emission line near a steep turnover cannot be reproduced straightforwardly by standard SED modelling codes such as \textsc{bagpipes}, and therefore will be modelled separately in Spectral modelling.

We adopted the non-parametric star formation history (SFH) of \citetsup{2019ApJ...876....3L} with $6$ bins in lookback time $t$. As in \citetsup{2023MNRAS.522.6236T}, the first two of these are spaced between $0 \, \mathrm{Myr} < t < 5 \, \mathrm{Myr}$ and $5 \, \mathrm{Myr} < t < 10 \, \mathrm{Myr}$, with logarithmic spacing for the $4$ remaining bins up to $z = 20$. We adopted the `bursty-continuity' prior for the SFR ratio between adjacent time bins\citesup{2022ApJ...927..170T}, having confirmed that consistent results are achieved for a more smoothly varying SFH (`standard continuity'). We varied the total stellar mass formed over $0 < M_* < 10^{15} \, \mathrm{M_\odot}$ and stellar metallicity across $0.001 \, \mathrm{Z_\odot} < Z_* < 1.5 \, \mathrm{Z_\odot}$, both with log-uniform priors. Our default model includes nebular emission, which \textsc{bagpipes} derives from \textsc{cloudy} models\citemet{2017RMxAA..53..385F}, where the nebular metallicity is fixed to the stellar one. The incident radiation field is set by the relevant SPS models, modulated by a freely varying ionisation parameter ($-3 < \log_{10} U < -0.5$). We included a flexible \citetsup{2000ApJ...539..718C} prescription to model the dust attenuation, as detailed in \citetm{2025MNRAS.536...27W} (see also ref. \citesup{2019MNRAS.483.2621C}). The PRISM spectrum was assumed to have a spectral resolution as derived for a uniformly illuminated micro-shutter, noting that fitting only to the continuum (while masking the observed line and break) the precise spectral resolution will not significantly impact the modelled observed spectrum. A first-order polynomial correction to the spectroscopic data\citesup{2019MNRAS.490..417C} was included to account for the mild discrepancy between NIRSpec and NIRCam (Synthetic photometry and path-loss corrections).

In our default model, we allow the redshift to vary within an interval of $\Delta z = 0.1$ centred on $z = 13.0$, so that we may explicitly marginalise over the uncertainty of the systemic redshift (further discussed in Spectral modelling). In addition, we considered a variation of the default model where we explicitly turn off nebular emission: this would be the case if the galaxy has a LyC escape fraction close to $f_\text{esc, LyC} = 100\%$. The resulting best-fitting models are shown in Supplementary Material \cref{fig:SED_models} and summarised in Supplementary Material \cref{tab:SED_models}.

Also shown in Supplementary Material \cref{fig:SED_models} is an alternative (`Balmer-break') model forced to be at lower redshift by centring the $\Delta z = 0.1$ interval on $z = 3.583$ instead, in which case the observed spectral break would correspond to a Balmer break and the (in this case not masked) observed emission line would be \OII (see Emission-line properties). When fitting this model, we include the MIRI/F770W photometric point. In agreement with \citetm{2024ApJ...964...71H} and \citetm{2024ApJ...970...31R}, we find this low-redshift solution produces a poor fit to the observed SED of \JGSzthirteenLA ($\chi^2 = 381.5$ over $15$ NIRCam and MIRI bands), whereas the data are reproduced much better by a model galaxy at $z \approx 13.0$ instead ($\chi^2 = 52.6$ in the default model).

Even so, the comparatively high $\chi^2$ (with $p = 4.5 \times 10^{-6}$ for the default model) indicates that photometric errors are potentially underestimated, or that the standard SPS models fall short of accurately describing the data. While the \textsc{forcepho} photometry circumvents the correlated noise between pixels in the mosaic images, it may still suffer from a degree of systematic uncertainty, including from any imperfections in the sky background subtraction. This effect is taken into account empirically by the aperture photometry (CIRC2), where the estimated uncertainty considers the scatter found in a number of randomly placed empty apertures, a detailed discussion of which can be found in \citetmet{2023ApJS..269...16R}. The true statistical errors therefore likely consist of a combination of the nominal uncertainty estimated as part of the \textsc{forcepho} modelling and CIRC2 aperture photometry (which covers approximately 70\% of the encircled energy of the F444W PSF). When we conservatively combine the estimated \textsc{forcepho} and CIRC2 photometric uncertainties (treating them as statistically independent) to better account for systematic effects, we instead find a more reasonable $\chi^2 = 23.7$ ($p = 0.070$).

Interestingly, we find the goodness of fit in the absence of nebular emission ($\chi^2 = 54.3$ or more conservatively $\chi^2 = 25.2$) to be essentially the same as in the default model, even if the inferred SFHs vary substantially between these models: the SFH in the model without nebular emission steadily increases, whereas the default model prefers an SFH that rises initially, but subsequently declines over the last $10 \, \mathrm{Myr}$. We conclude this downturn in SFH must be artificial, and only preferred by the default model in order to reproduce the very steep UV slope of $\beta_\text{UV} \lesssim -2.7$ independently measured both by NIRCam and NIRSpec (see UV magnitude and slope), as this requires reducing the number of very young ($<10 \, \mathrm{Myr}$) OB-type stars, which produce a large number of ionising photons at fixed UV luminosity\citem{2022ARA&A..60..455E}, thus minimising the contribution of nebular-continuum emission (which typically has $\beta_\text{UV} \approx -2$; refs. \citem{2024MNRAS.534..523C}$^,$\citesup{2024arXiv241114532S}) that drives the UV slope upwards\citesup{2017ApJ...840...44B, 2024MNRAS.531..997C, 2024MNRAS.529.4087T, 2024arXiv240410751A}.

The tendency towards a high LyC escape fraction is independently confirmed by \textsc{beagle} (BayEsian Analysis of GaLaxy sEds\citemet{2016MNRAS.462.1415C}) modelling with varying $f_\text{esc, LyC}$, largely as described in \citetm{2023NatAs...7..622C}. Briefly, we fitted the NIRSpec/PRISM spectrum only, again masking the spectral region around \Lya, $1150 \, \Angstrom < \lambda_\text{emit} < 1450 \, \Angstrom$. We assumed a constant SFH and a \citetsup{2003PASP..115..763C} IMF with stellar mass upper limit of $300 \, \mathrm{M_\odot}$. Reassuringly, the best-fitting parameters (Supplementary Material \cref{tab:SED_models}) are in good agreement with the \textsc{bagpipes} results despite different approaches, particularly considering the absence of normalisation to the NIRCam photometry in the \textsc{beagle} fit. The \textsc{beagle} fit yields a posterior $f_\text{esc, LyC} = 0.81_{-0.32}^{+0.14}$ (median and \nth{16}-\nth{84} percentiles).


\bibliographysup{GS-z13-1-LA}

\end{document}